\documentclass[smallextended]{svjour3}       % onecolumn (second format)

\usepackage[table,xcdraw,dvipsnames]{xcolor} %Enables colors, so that we can use it for review
\usepackage[utf8]{inputenc} %UTF 8 support
\usepackage[T1]{fontenc}
\usepackage{graphicx}
\usepackage{glossaries}
\usepackage{soul} %Allows text highlighting, so that we can use it for review
\usepackage[bookmarksdepth=2]{hyperref} %Makes crossrefs clickable
\usepackage[misc,geometry]{ifsym}  %Letter graphics for corresponding author
\usepackage{float} %Enables 'H' float
\usepackage[sort&compress,square,comma,numbers]{natbib}
\usepackage[many]{tcolorbox} %Used for summary boxes
\usepackage{enumitem} %Flexible enumerate, itemize, etc.
\usepackage{ifthen} %ifthenelse command (used by text commenting feature)
\usepackage{amssymb} %symbols (used by text commenting feature)
\usepackage{booktabs}
\usepackage{graphicx}
\usepackage{makecell}
\usepackage{amsmath}
\usepackage[export]{adjustbox}
\usepackage{ragged2e} %/justifying
\usepackage{relsize} %/smaller
\usepackage{multirow} %multicolumn support
\usepackage{pbox} %used in tables
\usepackage{balance} %balance refs across columns
\usepackage{moresize} %/ssmall, which is between \scriptsize and \tiny
\usepackage{xspace}

\newtcolorbox{mybox}[2][]{
top=0.15in,left=4pt,right=4pt,bottom=4pt,
fonttitle=\bfseries,
colbacktitle=gray,
colback=gray!5,
colframe=gray!40!black,
enhanced,
attach boxed title to top left={xshift=1.5em,yshift=-\tcboxedtitleheight/2},
boxed title style={size=small},
drop shadow={black!50!white},
title=#2,#1}

\newcommand{\countobservations}{
    \def \countobservations{1}
}
\newcounter{observation}
\newenvironment{observation}[1]
{   \refstepcounter{observation}
    \noindent \textbf{Observation~\theobservation) \textit{#1}}
}

\countobservations

\newcommand{\countimplications}{
    \def \countimplications{1}
}
\newcounter{implication}
\countimplications

\newboolean{showcomments}
\setboolean{showcomments}{true}

\ifthenelse{\boolean{showcomments}}
{\newcommand{\nbc}[3]{
 {\colorbox{#3}{\bfseries\sffamily\scriptsize\textcolor{white}{#1}}}
 {\textcolor{#3}{\sf\small$\blacktriangleright$\textit{#2}$\blacktriangleleft$}}
 }
}
{\newcommand{\nbc}[3]{}
 }

\newcommand{\smalltt}[1]{\ifmmode{\mbox{\smaller\texttt{#1}}}\else{\smaller\tt #1}\fi}
\newcommand{\code}[1]{\smalltt{#1}}

\newcolumntype{L}[1]{>{\raggedright\let\newline\\\arraybackslash\hspace{0pt}}m{#1}}
\newcolumntype{C}[1]{>{\centering\let\newline\\\arraybackslash\hspace{0pt}}m{#1}}
\newcolumntype{R}[1]{>{\raggedleft\let\newline\\\arraybackslash\hspace{0pt}}m{#1}}

\newcommand{\diffvalue}{$\mathit{diff_{value}}$\xspace}

\usepackage{soul}
\usepackage{moresize}
\usepackage{listings}
\begin{document}

	%Title
	\title{Using Knowledge Units of Programming Languages to Recommend Reviewers for Pull Requests: An Empirical Study}
	
	%Short title
	\titlerunning{Using KUs of PLs to Recommend Reviewers for PRs: An Empirical Study}
	
	%Authorship
	\author{Md Ahasanuzzaman \and Gustavo A. Oliva \and Ahmed E. Hassan}

	\institute{
		\Letter \space Md Ahasanuzzaman, Gustavo A. Oliva, and Ahmed E. Hassan \at
		Software Analysis and Intelligence Lab (SAIL), School of Computing \\
		Queen's University, Kingston, Ontario, Canada\\    
		\email{\{ma87,gustavo,ahmed\}@cs.queensu.ca}
	}

	\date{Received: date / Accepted: date}

	\maketitle

	\begin{abstract}
		\justifying{Code review is a key element of quality assurance in software development. Determining the right reviewer for a given code change requires understanding the characteristics of the changed code, identifying the skills of each potential reviewer (expertise profile), and finding a good match between the two. To facilitate this task, we design a code reviewer recommender that operates on the knowledge units (KUs) of a programming language. We define a KU as a cohesive set of key capabilities that are offered by one or more building blocks of a given programming language. We operationalize our KUs using certification exams for the Java programming language. We detect KUs from 10 actively maintained Java projects from GitHub, spanning 290K commits and 65K pull requests (PRs). Next, we generate developer expertise profiles based on the detected KUs. Finally, these KU-based expertise profiles are used to build a code reviewer recommender (KUREC). The key assumption of KUREC is that the code reviewers of a given PR should be experts in the KUs that appear in the changed files of that PR. In RQ1, we compare KUREC's performance to that of four baseline recommenders: (i) a commit-frequency-based recommender (CF), (ii) a review-frequency-based recommender (RF), (iii) a modification-expertise-based recommender (ER), and (iv) a review-history-based recommender (CHREV). We observe that KUREC performs as well as the top-performing baseline recommender (RF). From a practical standpoint, we highlight that KUREC's performance is more stable (lower interquartile range) than that of RF, thus making it more consistent and potentially more trustworthy. Next, in RQ2 we design three new recommenders by combining KUREC with our baseline recommenders. These new combined recommenders outperform both KUREC and the individual baselines. Finally, in RQ3 we evaluate how reasonable the recommendations from KUREC and the combined recommenders are when those deviate from the ground truth. KUREC is the recommender with the highest percentage of reasonable recommendations (63.4\%). One of our combined recommenders (AD\_FREQ) strikes the best balance between sticking to the ground truth (best recommender from RQ2) and issuing reasonable recommendations when those deviate from that ground truth (59.4\% reasonable recommendations, third best in this RQ). Taking together the results from all RQs, we conclude that KUREC and AD\_FREQ are overall superior to the baseline recommenders that we studied. Future work in the area should thus (i) consider KU-based recommenders as baselines and (ii) experiment with combined recommenders.}

%Draft abstract for submission system - must be 150 words only (double-check):
%In this paper, we introduce the notion of knowledge units (KUs) of programming languages to represent developer's expertise profiles and demonstrate how KU-based expertise profiles can be useful for recommending code reviewers in pull requests (PRs). We define a KU as a cohesive set of key capabilities that are offered by one or more building blocks of a given programming language. We operationalize our KUs via certification exams of the Java programming language (e.g., Oracle Java SE and EE exams). In this study, we analyze and detect KUs from 10 actively maintained Java projects from GitHub, spanning 290K commits and 65K PRs. We generate expertise profiles with the detected KUs for every developer of the studied projects. We use such KU-based expertise profiles to build a code reviewer recommender (KUREC). We compare KUREC's performance to that of four baseline recommenders. We observe that the KUREC has a more stable performance than the top-performing baseline recommenders and outperforms the remaining three baselines. Combining the KUREC with all other baseline recommenders in a straightforward manner even generates a higher-performing recommender.
	\end{abstract}

    \newcommand\PrelimStudy{Do KUs provide a new lens to study developers' expertise?}
    \newcommand\RQOne{How accurately can KUREC recommend code reviewers in pull requests?}
    \newcommand\RQTwo{Can KUREC be made more accurate by combining it with existing recommenders?}
	\newcommand\RQThree{How reasonable are the recommendations made by KUREC?}

    %Sections
    \section{Introduction}
\label{sec:Intro}

%enables finding expert developers on a given topic as well as sets of developers with similar expertise. 

Code review has a prominent role in modern software development. Indeed, code review systems are baked into the most popular software development platforms (a.k.a, forges), including GitHub, GitLab, and BitBucket. 

Code review comprises ``a record of the interaction between the owner of a change and reviewers of the change, including comments on the code and signoffs from reviewers''~\citep{chrev_automatic_review_recommend_ASE_2015}. The quality of a code review thus inherently depends on the selection of the reviewer. Yet, finding the right reviewer for a set of code changes is not always a trivial task. This is particularly true in the context of large-scale, distributed software development ~\citep{bird2013expectations,rigby2013}.

Mapping different expertises to individual developers is a key requirement for effective code review. Recent research studies have attempted to develop approaches to detect experts in specific topics~\citep{montandon2019identifying} and accurately represent expertise~\citep{expert_represent_ICSE_2021}. In this paper, we focus on a key facet of expertise, which is \textit{programming language (PL) expertise}. Our rationale is that a piece of code involving concurrency is suitable to be reviewed by someone who has demonstrated experience in dealing with such type of code.

\begin{figure}[!htbp]
    \centering
    \includegraphics[width=1.0\linewidth]{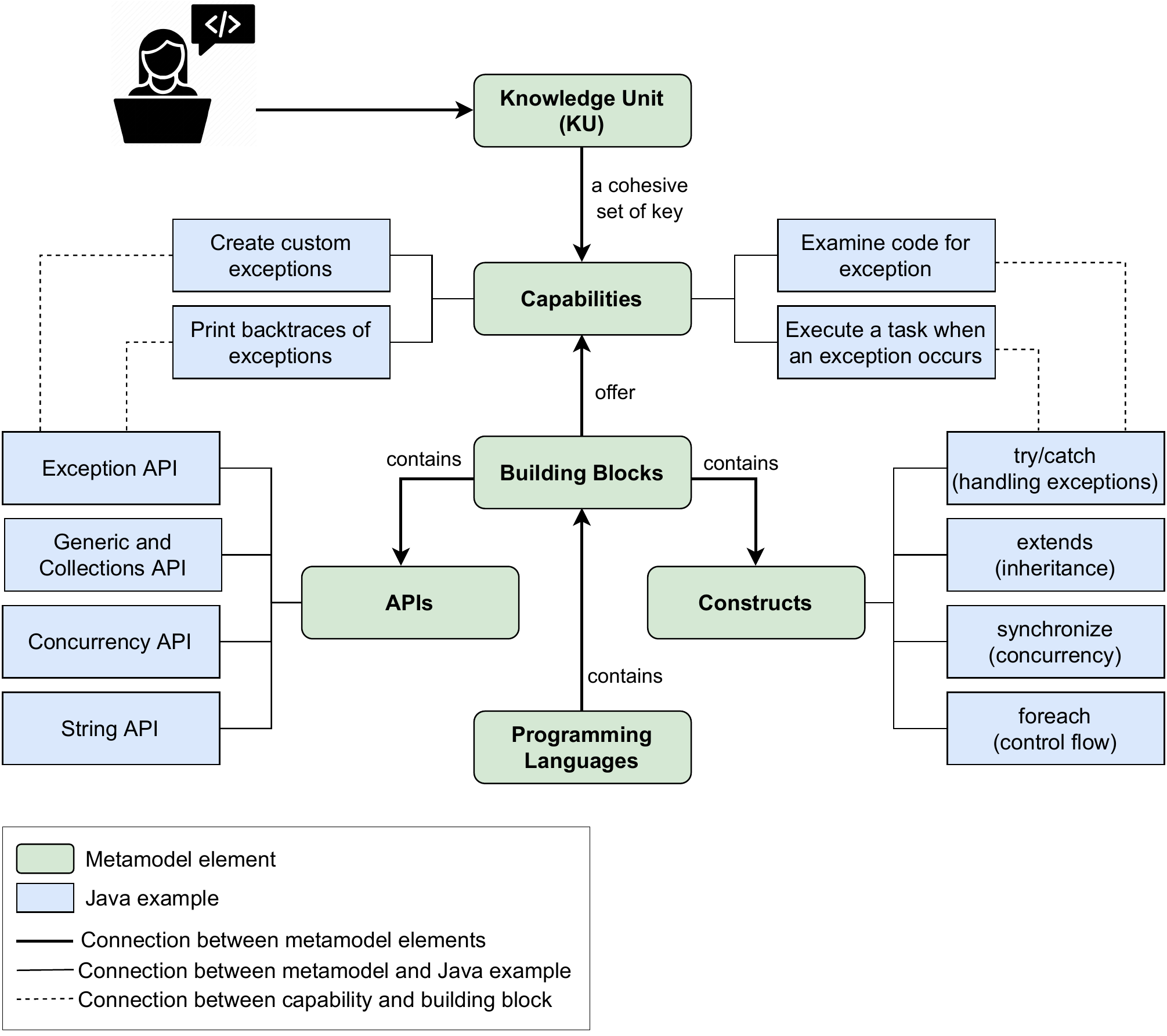}
    \caption{Our metamodel for knowledge units (KUs).}
    \label{fig:ku_metamodel}
\end{figure}

In order to capture the programming language expertise of developers, we introduce the notion of \textit{Knowledge Units (KUs)} of programming languages. We define a KU as \textit{a cohesive set of key capabilities that are offered by one or more building blocks of a given programming language} (see Figure \ref{fig:ku_metamodel}). For instance, some key capabilities offered by the \textit{Exception} building block of the Java programming language include (i) creating custom exception classes and (ii) printing the stack traces of exceptions. To operationalize the definition of KUs for Java, we analyze the topics covered in Java certification exams. Our rationale is that exam topics are precisely aimed at evaluating the degree of one's expertise in applying the key capabilities that are offered by the building blocks of Java. To determine developers' expertise based on KUs, we detect the KUs that are present in the Java files that developers change in their commits. 

This paper reports an empirical study on using the knowledge units of the Java programming language to recommend reviewers for GitHub pull requests (PRs). We start with a preliminary study in which we evaluate whether our operationalization of Java KUs can reveal software developers with different programming language expertise. To conduct this study, we detected and analyzed KUs from 10 real-world actively maintained Java projects from GitHub, spanning a total of 290k commits. Next, we applied the k-means algorithm to cluster developers based on the KUs that are present in their changed files. Running the algorithm resulted in the identification 71 different developer clusters of varying sizes. A closer inspection of the clusters indeed revealed developers with different KU expertise profiles.

Given the encouraging results from the preliminary study, we proceeded to build a KU-based reviewer recommendation systems (KUREC). KUREC works as follows. Given a certain PR, we construct a KU-based \textit{development expertise profile} and a KU-based \textit{review expertise profile} for every developer. More specifically, these profiles are created based on the KUs that a developer encountered during his prior code changes and code reviews. Next, a PR-specific \textit{expertise score} is computed for every developer based on how well the KUs that appear in the changed files of the PR in question match the two aforementioned expertise profiles. This score aims to capture whether a candidate reviewer has proven experience with the KUs that are present in the changed files of the PR. Finally, the score gets a bonus depending on how recent the candidate's prior code changes and code reviews are. The intuition is that someone who has worked with concurrent code last week is probably more suitable to review a PR involving concurrent code than someone who last worked with concurrent code 6 months ago. Finally, we simply rank candidate reviewers based on their expertise score and recommend the top ones. With KUREC in hand, we address the following three research questions: 

\smallskip \noindent \textbf{RQ1: \RQOne} We compare the performance of KUREC to that of four other baseline recommenders, namely: (i) a commit-frequency-based recommender (CF)~\citep{dev_interaction_exp_MSR_2013}, (ii) a review-frequency-based recommender (RF)~\citep{who_should_review_pull_request}, (iii) a modification-expertise-based recommender (ER)~\citep{exp_recom_cscw, code_rev_recom_empirical_ASE_2016}, and (iv) a review-history-based recommender (CHREV)~\citep{chrev_automatic_review_recommend_ASE_2015}. Recommender performance is computed using two measures: top-k accuracy and mean average precision (MAP). Our results indicate that:

\smallskip \noindent \textit{KUREC performs as well as the top-performing baseline recommender (RF) and outperforms the remaining three baselines. We also note that KUREC has a more stable performance (smaller interquartile range) compared to RF, which is a desired property in practice~\citep{stability_recommendation_system}.}

\smallskip \noindent \textbf{RQ2: \RQTwo} Inspired by our prior work on adaptive heuristics~\citep{malik2008supporting}, we combine KUREC with all other baseline recommenders to create three brand-new recommenders: a frequency-based one called AD\_FREQ (given a PR, AD\_FREQ triggers the baseline recommender that performed the best most of the time), a recency-based one called AD\_REC (given a PR, AD\_REC triggers the baseline recommender that performed the best in the prior PR), and a frequency/recency hybrid one called AD\_HYBRID (given a PR, AD\_HYBRID triggers the baseline recommender that performed the best in the prior 10 PRs). We refer to AD\_FREQ, AD\_REC, and AD\_HYBRID as combined recommenders. Our results indicate that:

\smallskip \noindent \textit{All three combined recommenders have similar performance. For top-k accuracy, the combined recommenders perform as well as the top baselines (RF and KUREC). For MAP, the combined recommenders outperform the top baselines. More specifically, combined recommenders achieve a median MAP that ranges from 0.56 to 0.57, which is 16.5\%--18.4\% higher than that of RF and 34\%--36.1\% higher than that of KUREC. Finally, the combined recommenders always outperform the baselines in both top-1 accuracy and MAP@1. We thus conclude that combining KUREC with other recommenders using a simple approach results in a more accurate recommender.}

\smallskip \noindent \textbf{RQ3: \RQThree} To evaluate KUREC in RQ1 and RQ2, we compared its recommendations to a ground truth. This ground truth corresponds to the list of actual developers who reviewed the PRs. The results indicated that sometimes KUREC's recommendations differ from such ground truth. We thus wonder whether the recommendations made by KUREC (and our other studied recommenders) would have been reasonable choices in practice when those differ from what happened in reality. We consider a recommendation to be \textit{reasonable} if the (top-1) recommended individual recently committed or submitted PRs that included the majority of the files from the PR in question. Our results indicate that:

\smallskip \noindent \textit{KUREC is the recommender with the highest percentage of reasonable recommendations (63.4\%). The AD\_FREQ combined recommender, in particular, strikes the best balance between sticking to the ground truth (best recommender from RQ2) and issuing reasonable recommendations when those deviate from that ground truth (59.4\%, third best recommender in this RQ).}

\smallskip Taking together the results from all RQs, we conclude that KUREC and AD\_FREQ are overall superior to the baseline recommenders that we studied. 

\smallskip \noindent \textbf{The main contribution of our paper are as follows:} (i) operationalizing the notion of knowledge units of programming languages using a systematic approach, (ii) defining a method to represent developers' programming language expertise using the knowledge units (KUs) of a programming language, and (iii) introducing a KU-based reviewer recommender (KUREC) and variations (e.g., AD\_FREQ) that are overall superior to the baseline recommenders that we studied. Future work in the area should thus consider KU-based recommenders as baselines and experiment with combined recommenders. A supplementary material package is provided online\footnote{\url{http://www.bit.ly/3bhSFux}. The contents will be made available on a public GitHub repository once the paper is accepted.}

\smallskip \noindent \textbf{Paper organization.} Section~\ref{sec:Knowledge_Unit} defines KUs and presents our approach to detect them. Section~\ref{sec:Data_Collection} describes our data collection approach. Section~\ref{sec:Preliminary_Study} reports our preliminary study in which we analyze the potential behind KUs. Section~\ref{sec:Main_study} presents the motivation, approach, and findings of our three research questions. Section~\ref{sec:Related_Work} discusses related work. Section~\ref{sec:Limitations_And_Threats} describes the threats to the validity of our findings. Finally, Section~\ref{sec:Conclusion} outlines our concluding remarks.

	\section{Knowledge Units (KUs)}
\label{sec:Knowledge_Unit}

\subsection{Definition}
\label{subsec:knowledge_definition}

Every programming language consists of \textit{building blocks} that developers use to produce code. In the case of Java programming language, we consider the building blocks to be the Java fundamental \textit{language constructs} (e.g., \texttt{if/else} statements, \texttt{try/catch} statements, and \texttt{synchronize} blocks) and \textit{APIs} (e.g., Exception API, Concurrency API, and Generic and Collections API). Each building block offers a \textit{set of capabilities}, which are essentially the things that a developer can do with that building block.

A Knowledge Unit (KU) is a cohesive set of key capabilities that are offered by one or more building blocks of a given programming language. Our definition includes the word ``key'' and our rationale is to avoid having KUs conceived around capabilities that are too specific or highly context-dependent. Figure~\ref{fig:ku_metamodel} illustrates key capabilities that are associated with the Exception API (left-hand side) and the \texttt{try/catch} language construct (right-hand side).

\subsection{Operational Definition}
\label{subsec:knowledge_operational_definition}

Certification exams of a programming language (e.g., Oracle Java SE and Java EE certification exams for the Java programming language) aim to determine the skills and knowledge of a developer in using the key capabilities offered by the building blocks of that language. Hence, we can say that \textit{certification exams capture the KUs of a programming language}. For instance, one of the topics covered in the Java SE 8 Programmer II certification exam is ``Concurrency''. The subtopics of ``Concurrency'' that are covered in the exam include: (i) create worker threads using \texttt{Runnable}, \texttt{Callable} and use an \texttt{ExecutorService} to concurrently execute tasks, (ii) use \texttt{synchronize} keyword and \texttt{java.util.concurrent.atomic} package to control the order of thread execution, (iii) use \texttt{java.util.concurrent} collections and classes including \texttt{cyclicBarrier} and \texttt{copyonWriteArrayList}, and (iv) use parallel fork/join framework. We interpret such a list of subtopics as the key capabilities that are offered by the ``Concurrency'' building block of the Java programming language. From this interpretation, we infer that Java has a ``Concurrency KU''.

In this paper, we operationalize our KUs via Oracle certification exams for the Java programming language. Oracle offers certification exams for different Java editions, such as Java Standard Edition (Java SE) and Java Enterprise Edition (Java EE). For Java SE, Oracle has two exam versions: (i) \textit{Oracle Certified Associate, Java SE 8 Programmer I Certification exam}~\citep{oracle_se_oap}, and (ii) \textit{Oracle Certified Professional, Java SE 8 Programmer II Certification exam }\citep{oracle_se_ocp}. For Java EE, Oracle offers only one version of the exam: \textit{Oracle Certified Professional, Java EE Application Developer Certification exam}~\citep{oracle_ee_prof}. In this study, we extract the list of KUs directly from the topics of the aforementioned certification exams. Each topic of each certification exam consists of a list of subtopics, and we interpret such subtopics as the key capabilities of the building blocks of the Java programming language. More specifically, to operationalize KUs we perform the following steps. First, we eliminate subtopics (from any exam) that are inappropriate for our research (i.e.,  subtopics that explain concepts and cannot be extracted automatically). Subtopics are then rearranged (both inside and across tests) so that they fall under the most specific topic possible. Finally, each topic is converted into a KU (one-to-one mapping), and its subtopics are interpreted as the KU's main capabilities. To mitigate subjectiveness bias, the first two authors collaborated on the entire mapping process and discussed how to handle corner cases. At the end of this process, we identified 28 KUs (Table~\ref{tab:topic_definition}). Appendix \ref{appendix:cert-exams} details the mapping process.

\begin{table*}[!htbp]
    \centering
    \caption{The identified Java programming language knowledge units (KUs).}
    \label{tab:topic_definition}
    \resizebox{\columnwidth}{!}{
    \begin{tabular}{p{3.2cm}p{13.8cm}}
        \toprule
        \multicolumn{1}{C{3cm}}{\textbf{Knowledge unit (KU)}} & \multicolumn{1}{C{14cm}}{\textbf{Definition}}                                                                                                                                                                                                                               \\ \midrule

        \textbf{[K1]} Data Type KU                                          &
        This KU describes the declaration and initialization of different types of variables (e.g., primitive type and parameterized type.)
        \\ \midrule

        \textbf{[K2]} Operator and Decision KU                              &
        This KU describes the usage of different Java operators (e.g., assignment, logical, and bit-wise operators) and conditional statements (e.g, if, if-else, and switch statements).
        \\ \midrule

        \textbf{[K3]} Array KU                                              &
        This KU describes declaration, instantiation, initialization and the usage of one-dimensional and multi-dimensional arrays.
        \\ \midrule

        \textbf{[K4]} Loop KU                                               &
        This KU describes the execution of a set of instruction-s/methods repeatedly using for, while, and do-while statements and the skipping and stopping of a repetitive execution of instructions and methods using continue and break statements.
        \\ \midrule

        \textbf{[K5]} Method and Encapsulation KU                           &
        This KU describes the creation of methods with parameters, the use of overloaded methods and constructors, the usage of constructor chaining, the creation of methods with variable length arguments, and the usage of different access modifiers. This KU also describes encapsulation mechanisms, such as creating a set and get method for controlling data access, generating immutable classes, and updating object type parameters of a method.
        \\ \midrule

        \textbf{[K6]} Inheritance KU                                        &
        This KU describes developing code with child class and parent class relationship, using polymorphism (e.g., developing code with overridden methods), creating abstract classes and interfaces, and accessing methods and fields of the parent's class.
        \\ \midrule

        \textbf{[K7]} Advanced Class Design KU                              &
        This KU describes developing code that uses the final keyword, creating inner classes including static inner classes, local classes, nested classes, and anonymous inner classes, using enumerated types including methods and constructors in an enum type, and developing code using @override annotator.
        \\ \midrule

        \textbf{[K8]} Generics and Collection KU                            &
        This KU describes the creation and usage of generic classes, usage of different types of interfaces (e.g., List Interface, Deque Interface, Map Interface, and Set Interface), and comparison of objects using interfaces (e.g., java.util.Comparator, and java.lang.Comparable).
        \\ \midrule

        \textbf{[K9]} Functional Interface KU                               &
        This KU describes developing code that uses different versions of defined functional interfaces (e.g., primitive, binary, and unary) and developing code with user-defined functional interfaces.
        \\ \midrule

        \textbf{[K10]} Stream API KU                                         &
        This KU describes developing code with lambda expressions and Stream APIs. This includes developing code to extract data from an object using peek() and map() methods, searching for data with search methods (e.g., findFirst) of the Stream classes, sorting a collection using Stream API, iterating code with foreach of Stream, and saving results to a collection using the collect method.
        \\ \midrule

        \textbf{[K11]} Exception KU                                          &
        This KU describes the creation of try-catch blocks, the usage of multiple catch blocks, the usage of try-with-resources statements, the invocation of methods throwing an exception, and the use of assertion for testing invariants.
        \\ \midrule

        \textbf{[K12]} Date time API KU                                      & This KU describes creating and managing date-based and time-based events using Instant, Period, Duration, and TemporalUnit, and working with dates and times across timezones and managing changes resulting from daylight savings, including Format date and times values. \\ \midrule

        \textbf{[K13]} IO KU                                                 &
        This KU describes reading and writing data from console and files and using basic Java input-output packages (e.g., java.io.package).
        \\ \midrule

        \textbf{[K14]} NIO KU                                                &
        This KU describes how to interact files and directories with the new non-blocking input/output API (e.g., using the Path interface to operate on file and directory paths. This KU also includes other file-related operations (e.g., read, delete, copy, move, and manage metadata of a file or directory).
        \\ \midrule

        \textbf{[K15]} String Processing KU                                  &
        This KU describes searching, parsing, replacing strings using regular expressions, and using string formatting.
        \\ \midrule

        \textbf{[K16]} Concurrency KU                                        &
        This KU describes many functionalities that are related to thread execution and parallel programming. Some of these functionalities include: creating worker threads using Runnable and Callable classes, using an ExecutorService to concurrently execute tasks, using synchronized keyword and java.util.concurrent.atomic package to control the order of thread execution, using java.util.concurrent  classes and using a parallel fork/join framework.
        \\ \midrule

        \textbf{[K17]} Database KU                                           &
        This KU describes the creation of database connection, submitting queries and reading results using the core JDBC API.
        \\ \midrule   

        \textbf{[K18]} Localization KU                                       &
        This KU describes reading and setting the locale (Oracle defines a locale as ``a specific geographical, political, or cultural region'') by using the Locale object, and building a resource bundle for each locale, and loading a resource bundle in an application.
        \\ \midrule

        \textbf{[K19]} Java Persistence KU                                    &
        This KU describes the usage of object/relational mapping facilities for managing relational data in Java applications. With this knowledge, developers can learn how to map, store, update and retrieve data from relational databases to Java objects and vice versa.                                                              \\ \midrule

        \textbf{[K20]} Enterprise Java Bean KU                               &
        This KU is the knowledge about managing server-side components that encapsulate the business logic of an application.                                                                                                                                                                                                               \\ \midrule

        \textbf{[K21]} Java Message Service API KU                           &
        This KU describes how to create, send, receive and read messages using reliable, asynchronous, and loosely coupled communication.                                                                                                                                                                                                   \\ \midrule

        \textbf{[K22]} SOAP Web Service KU                                   &
        This KU describes how to create and use the Simple Object Access Protocol for sending and receiving requests and responses across the Internet using JAX-WS and JAXB APIs.                                                                                                                                                          \\ \midrule

        \textbf{[K23]} Servlet KU                                            &
        This KU describes handling HTTP requests, parameters, and cookies and how to process them on the server sites with appropriate responses.                                                                                                                                                                                           \\ \midrule

        \textbf{[K24]} Java REST API KU                                      &
        This KU describes creating web s services and clients according to the Representational State Transfer architectural pattern using JAX-RS APIs.                                                                                                                                                                                     \\ \midrule

        \textbf{[K25]} Websocket KU                                          &
        This KU describes how to create and handle bi-directional, full-duplex, and real-time communication between the server and the web browser.                                                                                                                                                                                         \\ \midrule

        \textbf{[K26]} Java Server Faces KU                                  &
        This KU describes building UI component-based and event-oriented web interfaces using the standard JavaServer Faces (JSF) APIs.                                                                                                                                                                                                     \\ \midrule

        \textbf{[K27]} Contexts and Dependency Injection (CDI) KU            &
        This KU describes how to manage the lifecycle of stateful components using domain-specific lifecycle contexts and type-safely inject components (services) into client objects.                                                                                                                                                     \\ \midrule

        \textbf{[K28]} Batch Processing KU                                   &
        This KU describes how to create and manage long-running jobs on schedule or on demand for performing on bulk-data, and without manual intervention.                                                                                                                                                                                 \\ \bottomrule
    \end{tabular}
    }
\end{table*}

\subsection{Detection of KUs}
\label{subsec:knowledge_detection}

We employ a static analysis approach to detect KUs from a given Java project. First, we parse all Java source files of all the commits of the studied projects using the Eclipse JDT framework \citep{eclipse_jdt}. The JDT framework generates an Abstract Syntax Tree (AST) for the source code and provides visitors for every element of such tree, including \textit{names}, \textit{types} (class, interfaces), \textit{expressions}, \textit{statements}, and \textit{declarations}.\footnote{The grammar employed by JDT can be seen at \url{https://github.com/eclipse/eclipse.jdt.core/blob/master/org.eclipse.jdt.core/grammar/java.g}}

To detect a KU, we refer to the set of key capabilities of that unit (Table~\ref{tab:ku-from-java-exams}). For instance, to detect the presence of the \textit{Generics and Collections KU} ([K8]), we use the visitors provided by JDT to detect the capabilities associated with that unit, such as [K8, C1] \textit{create and use a generic class} and [K8, C2] \textit{create and use} \code{ArrayList}, \code{TreeSet}, \code{TreeMap}, \textit{and} \code{ArrayDeque} \textit{objects}. For instance, using type visitors, we can determine whether a given class is a generic class ([K8, C1]). Similarly, using statement visitors, we can find instantiations of an ArrayList ([K8, C2]).

During our code analysis, we also collect binding information of methods, classes, and variables (e.g., \code{org.eclipse.jdt.core.dom.IVariableBinding} resolves variable binding). In our study, we discard type bindings to third-party libraries because our focus is on studying the KUs that are associated with the source code that are written by the developers of the studied subject systems. Finally, we detect the presence of a KU if we can identify the usage of any of the key capabilities associated with that KU in a Java file.

We count the occurrences of KUs per Java source file. For example, one of the capabilities associated with the \textit{Inheritance KU} ([K6]) is to create abstract classes ([K6,C4]).  Hence, if we find a declaration of an \texttt{abstract} class in a given Java file \code{X.java} belonging to a subject system \textit{S}, then the counter for \textit{Inheritance KU} that is associated with file \texttt{X.java} is incremented by one. In other words, for each file, we produce a vector \textit{V} containing the counts of each KU in that file.

%\ahmed{associated capabilities --> fix everytwhere} \ahsan{fixed} 
%goliva{the 'associated with' is a transitive verb and should NOT be split. See this very reliable doc for an example https://media.defense.gov/2023/Mar/14/2003178390/-1/-1/0/CSI_Zero_Trust_User_Pillar_v1.1.PDF -> ``The user pillar expands and refines the capabilities associated with FICAM framework to address the enhanced threat to identity, credentials, and access management.''}
    \section{Data Collection}
\label{sec:Data_Collection}

\begin{figure}[!t]
    \centering
    \includegraphics[width=0.9\textwidth]{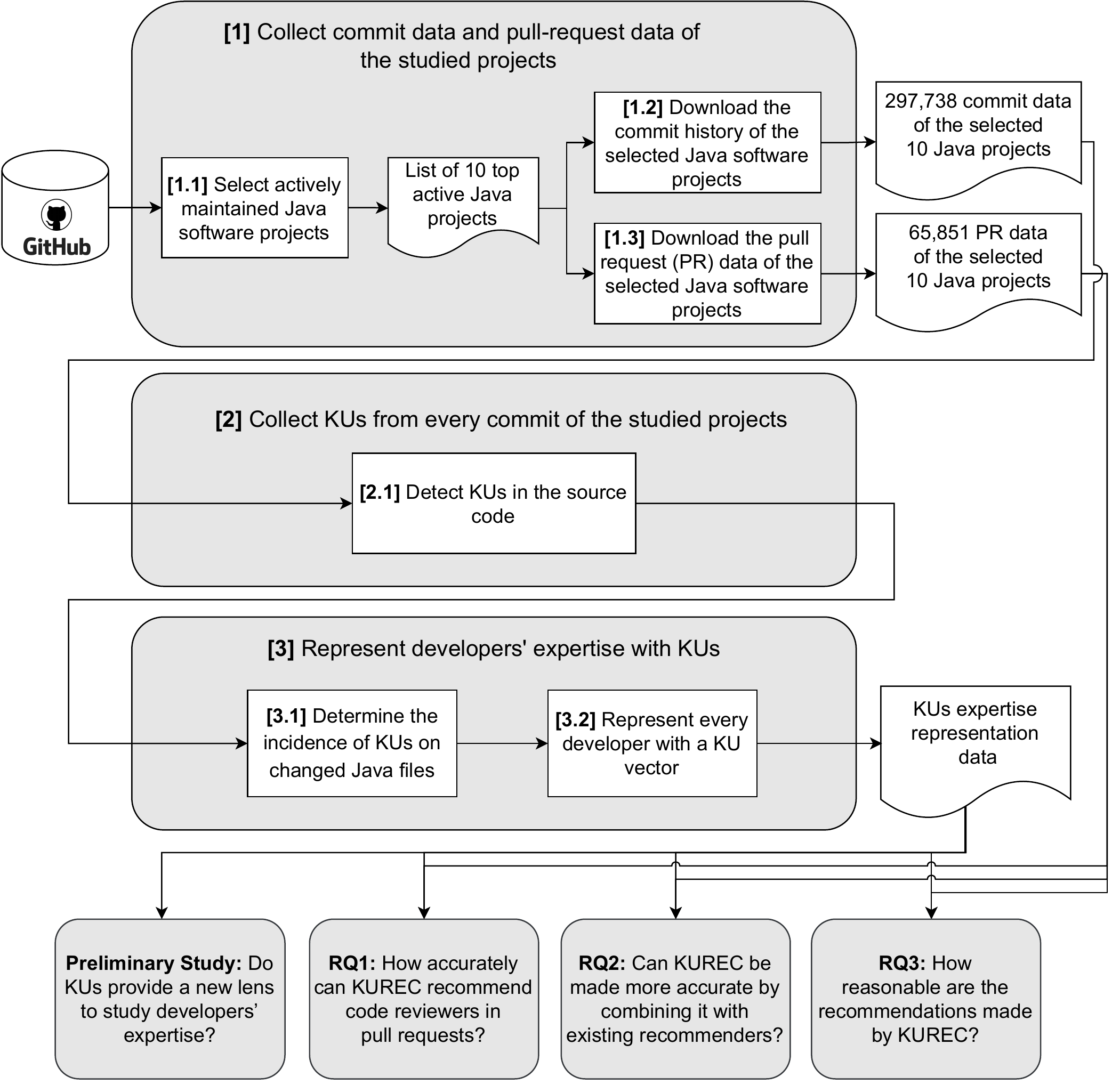}
    \caption{An overview of our data collection process.} 
    \label{fig:data_collection}
\end{figure}

Our data collection process contains three main steps. First, we retrieve the source code and pull-request (PR) data from real-world, active Java software projects. We refer to these projects as our \textit{studied projects}. Next, we determine the incidence of KUs in the source code of all commits of our studied projects. Finally, we represent developers' expertise with KUs. Figure~\ref{fig:data_collection} presents an overview of our data collection process. We describe each step in more detail below.

\begin{itemize}[wide = 0pt, label=$\bullet$]
    \item \textit{Step 1) Collect commit data and pull-request data of the studied projects:}
          \begin{itemize}[wide = 0pt, itemsep = 3pt, topsep=3pt, listparindent=\parindent]
              \item \textit{Step 1.1) Select actively maintained Java software projects.} In our study, we focus on Java projects hosted on GitHub and that are both popular and actively maintained. However, there exist toy projects, personal projects, academic projects, and websites on GitHub that fulfil those criteria~\citep{promise_perils_GitHub_MSR_2014}. Hence, in order to select real-world software projects, we use the dataset of \textit{engineered projects} made available by \citet{engineered_project_github_EMSE_2017}. More specifically, we first select the projects from the list that are tagged with the Java language. Next, to study popular and active projects, we filter out projects that have less than 100 stars. We also filter out those that did not receive commits in the past year (relative to our data collection date, which is January 30th, 2022). Then, we sort the list of projects in decreasing order of the \textit{average number of commits per month}. Finally, we select the top-10 projects from this list.

              \item \textit{Step 1.2) Download the commit history of the selected Java software projects.} To download the commit history of our studied software projects, we \texttt{git clone} the repository of each project. The list of projects and their associated GitHub repository links can be found in our supplementary material package. Then, we use PyDriller~\citep{pydriller_MSR_2018} to collect all the commit information from the projects, such as commit hash id, commit author name, commit author-date, commit message, and list of modified files in a commit. To retrieve the source code (snapshot) of a particular commit of a project, we use the command \texttt{git checkout commit\_hash}, where \texttt{commit\_hash} is the commit id.

              \item \textit{Step 1.3) Download the pull request (PR) data of the selected Java software projects.} To download the PR data of the studied projects, we use the GitHub Rest API\footnote{\url{https://docs.github.com/en/rest/reference}}. For example, to collect all the PRs of the Apache HBase project, we issue a \texttt{GET} request to \texttt{https://api.github.com/repos/apache/hbase/\\pulls?state=all\&per\_page=100}. We also use the GitHub API to collect review comments, discussion comments, changed files, and reviewer information for every PR. We use PRs and their associated metadata in RQ1 and RQ2.
                    %The list of all requests of GitHub API for collecting all PR information can be found in our supplementary material.
          \end{itemize}

    \item \textit{Step 2) Collect KUs from every commit of the studied projects:}

          \begin{itemize}[wide = 0pt, itemsep = 3pt, topsep=3pt, listparindent=\parindent]
              \item \textit{Step 2.1) Detecting KUs in the source code.} We detect the KUs in each Java file of every commit (i.e., every snapshot) of the entire history of our studied projects. To detect KUs, we employ our custom detector built on top of Eclipse JDT (c.f., Section~\ref{subsec:knowledge_detection}).
          \end{itemize}

    \item \textit{Step 3) Represent developers' expertise with KUs:}

          \begin{itemize}[wide = 0pt, itemsep = 3pt, topsep=3pt, listparindent=\parindent]
              \item \textit{Step 3.1) Determine the incidence of KUs on changed Java files.} First, we identify all the Java files that are changed in every commit by filtering in those files with a \texttt{.java} extension. Then, we collect the commit's author name (i.e., the developer who actually performed the code changes), such that we can map changed Java files to developers. Next, we count the occurrences of KUs in each of the changed Java files of every commit. \textbf{This count represents the KU expertise gained by the commit's author from having changed those Java files.} Our rationale for counting KUs from the entire file (instead of a diff) is that a developer often needs to understand the entire Java file in order to be able to modify it even a small portion of it. Finally, for every developer, we sum the count of the occurrences for every KU that is identified in each changed file of the developer (see Figure \ref{fig:ku_representation_process}). 

              \item \textit{Step 3.2) Represent every developer with a KU vector.} We represent the expertise of every developer with a 28-position KU vector. Each position corresponds to a different KU (Table~\ref{tab:topic_definition}). In each vector position, we store the ratio of the incidence of a KU. We calculate a developer's ratio of the incidence of a KU by dividing the sum of the occurrences of a KU detected from the changed files of a developer to the total occurrences of the KU detected from all the changed files (see Figure \ref{fig:ku_representation_process}b - for simplicity, we only show three KUs). At the end, we have a KU vector for every developer. This KU vector captures a developer's Java KU expertise. We henceforth refer to this vector as $P_{ku}(D)$, where $D$ is some studied developer.
              
              \begin{figure}[!t]
                \centering
                \includegraphics[width=0.8\textwidth]{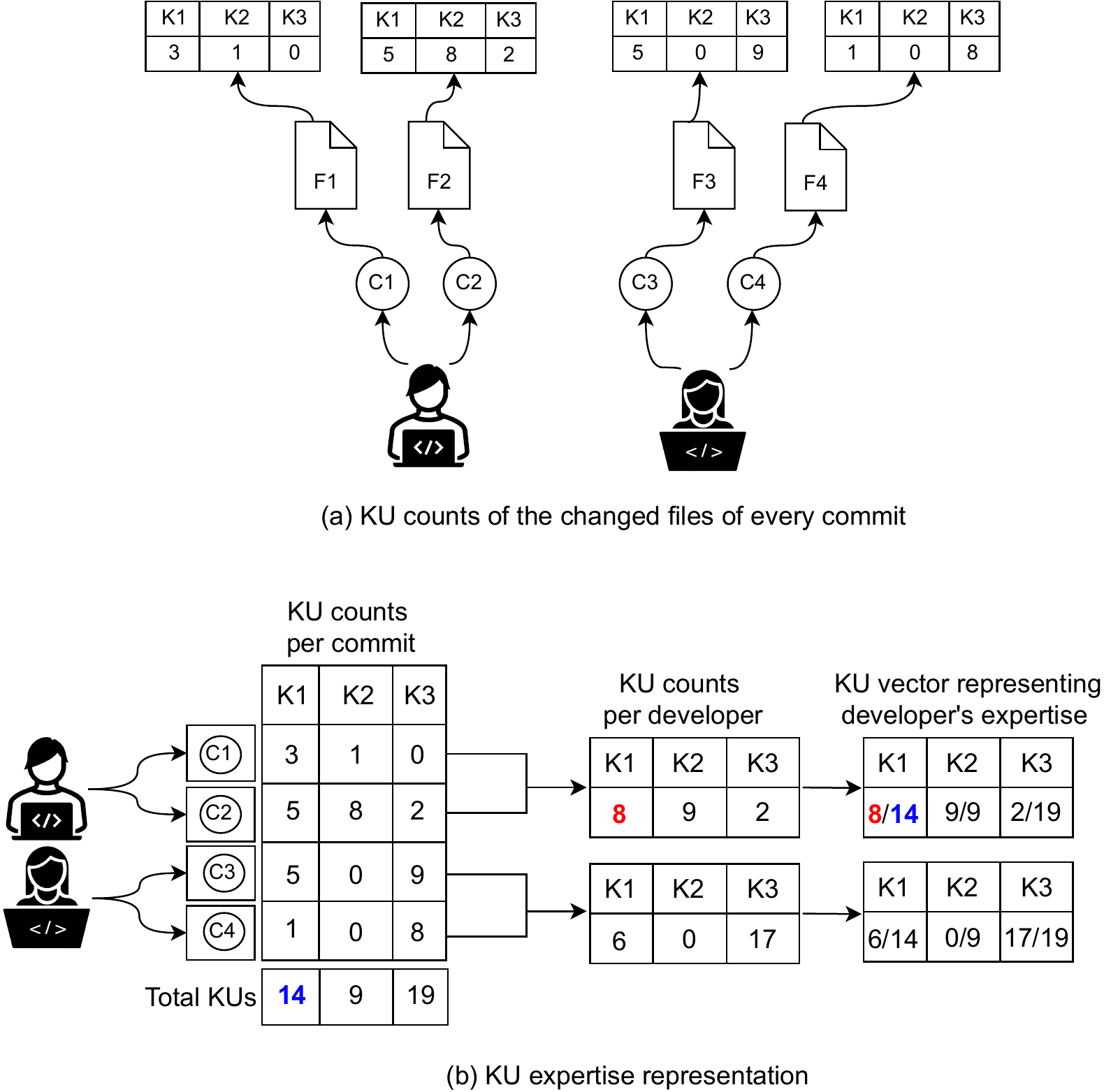}
                \caption{Representing developers' expertise with KUs. Here, $K_{i}$ denotes $i_{th}$ knowledge unit, $C_{j}$ denotes $j_{th}$ commit and $F_{j}$ denotes a changed file in $C_{j}$.} 
                \label{fig:ku_representation_process}
            \end{figure}
        \end{itemize}
\end{itemize}

\begin{footnotesize}
    \begin{mybox}{Summary}
        \begin{itemize}[itemsep = 3pt, label=\textbullet, topsep = 0pt, wide = 0pt]
            \item Data collection date: January 20th, 2022
            \item Data source: GitHub
            \item Number of studied Java projects: 10
            \item Number of studied commits: 297,738
            \item Number of pull requests: 65,851
            \item Pieces of data collected: java source code and pull requests 
            \item Datasets produced: developer expertise representation with KUs
        \end{itemize}
    \end{mybox}
\end{footnotesize}
    \section{Preliminary Study: \PrelimStudy}
\label{sec:Preliminary_Study}

\noindent \textbf{Motivation.} Our KU-based expertise profile is derived from the key capabilities of the Java building blocks. In this study, we want to group developers according to their KU-based expertise profiles and study such groups of developers. Finding several groups with unique KU profiles can indicate that KUs might be useful for code reviewer recommendations.

\smallskip \noindent \textbf{Approach.} We first cluster developers based on their KU-based expertise profile. Next, we study the dispersion of clusters' size. In the following, we detail each of the steps that we follow for this study.

\begin{itemize}[itemsep=3pt, wide = 0pt, topsep=1pt, label=$\bullet$, listparindent=\parindent]
	
	\item \textit{Step 1) Cluster developers based on their KU-based expertise profiles.} We take the KU-based expertise profile $P_{ku}(D)$ of every developer $D$ across all studied projects and glue them together to build a matrix $P_{ku}$ (the procedure for building $P_{ku}(D)$ is described in the step 3 of Section~\ref{sec:Data_Collection}). To cluster developers according to their KU-based expertise profile, we apply the k-means clustering algorithm to a dimentionality-reduced version of $P_{ku}$. K-means is one of the most popular clustering algorithms and has been widely used in software engineering research \citep{soft_fault_k_means, yoon_kmeans_soft_cluster, kmeans_software_cluster, cross_project_kmeans}. We detail the sub-steps of the clustering procedure below.
	
	\begin{itemize}[itemsep=3pt, wide = 0pt, topsep=1pt, listparindent=\parindent]

		\item \textit{Step 1a) Reduce the dimensionality $P_{ku}$.} Due to \textit{the curse of dimensionality}, many clustering algorithms, including the k-means algorithm,  perform poorly on high dimensional data~\citep{parson_dimensionality}. In simple terms, as the number of dimensions in a dataset increases, distance measures become less meaningful, since the distance between any pair of points in a high-dimensional space is almost the same over a wide range of data distributions and distance functions~\citep{Li_dimensionality_curse}. To overcome the dimensionality issue, a dimensionality reduction technique (e.g., Principal Component Analysis -- PCA) can be applied to the data prior to running the clustering algorithm \citep{kmeans_discussion, goebl2014finding}. Since PCA has been proven beneficial for the k-means algorithm~\citep{ding_pca_k_means}, we reduce the dimensionality of our data by applying PCA prior to running k-means.

		To apply PCA, we use the \texttt{prcomp}\footnote{\url{https://www.rdocumentation.org/packages/stats/versions/3.6.2/topics/prcomp}} native function of the R language. PCA transforms the data to a new coordinate plane, and each co-coordinate of that plane is called a principal component. PCA generates as many principal components as the number of dimensions of the data. The first principal component has the largest possible variance in the data, and each successor component has as much of the remaining variability as possible. Similarly to prior studies~\citep{xibilia2020soft, ghotra_pca_MSR_2017, kondo2019impact}, we reduce data dimensionality by selecting the first-n PCA components that account for at least 95\% of the explained variance of the data.

		\item \textit{Step 1b) Identify optimal number of clusters.} The k-means algorithm takes the number of clusters as an input parameter (\textit{K}). To identify the optimal number of clusters, we analyze the \textit{silhouette} coefficient~\citep{rousseeuw1987silhouettes}. The silhouette coefficient is a popular measure that has been widely used in prior studies~\citep{nidheesh2020hierarchical, cluster_mobile_apps_ESEM_2016, clustering_text_lda_ICSE_2013} for determining the optimal number of clusters. The silhouette coefficient indicates how similar an object (i.e., developer) is to its own cluster (cohesion) in comparison to other clusters (separation). The silhouette coefficient ranges from -1.0 to 1.0. A coefficient close to one indicates that the object well-matches its own cluster and, simultaneously, poorly-matches its neighboring clusters. If most objects have a high coefficient, then the clustering configuration is deemed appropriate. 
		
		We run k-means for \textit{K} values ranging from 2 to 100 and calculate the \textit{median silhouette} for each \textit{K} value. Our rationale for using the median in lieu of the mean is because the former is more robust to outliers (e.g., a clustering with only few very badly clustered objects should still be considered a good clustering overall)~\citep{rousseeuw1987silhouettes}. Next, we select the largest \textit{K} among those that yield a clustering with a median silhouette of 0.90 or higher. We avoid taking the \textit{K} that maximizes the median silhouette because it might be much lower than some other \textit{K} with an almost identical median silhouette. Our approach thus prioritizes finding more clusters instead of obtaining just marginally better clusterings (while still ensuring that the obtained clustering are appropriate due to the high threshold for median silhouette).
		
		%Such a threshold is high (i.e., threshold value is close to 1)  which can ensure that the clustering is well-formed.  do not select the K that maximizes the median silhouette value. Rather we use a high threshold because the maximum silhouette value could end up a very small number of clusters (e.g., median silhouette = 1.0 can find only 2 clusters (K=2)) while the high silhouette value (e.g., 0.90) not only could ensure relatively good clustering but also could find more clusters. Identifying diverse groups of developers (i.e., maximizes the number of clusters) while ensuring a good clustering overall is the expected goal for our study.
		
		\item \textit{Step 1c) Create clusters of developers.} To cluster developers using the k-means algorithm, we employ the native \texttt{kmeans} function of R language\footnote{\url{https://www.rdocumentation.org/packages/stats/versions/3.6.2/topics/kmeans}}. The input dataset corresponds to the selected PCA components in Step 1a. The number of clusters is set based on the results of Step 1b. Finally, we cluster developers using the k-means algorithm and visualize the clusters using the \texttt{fviz\_cluster} function of the \texttt{factoextra} R package~\citep{facto_extra}.
		
	\end{itemize}

    \item \textit{Step 2) Characterize developer clusters using KU-based expertise profiles.} To characterize developer clusters using KU-based expertise profiles, we follow two steps. First, we calculate a \diffvalue, which we define as the difference between the median value of a given KU within a cluster and the median value of the given KU across all the clusters (i.e., the entire dataset). The \diffvalue thus indicates how the median value of a KU within a cluster compares to the median value of that KU across all clusters. The intuition behind \diffvalue is to determine whether the values of a given KU $K$ in $C$ are either high or low (in relative terms). Formally, we define \diffvalue as follows: 

	\vspace{-2ex}
    \begin{align*}
        \mathit{diff_{value}(P_{ku},C,K)} = median(Subset(P_{ku},C,K)) \\ - \ median(Subset(P_{ku},K))
    \end{align*}
    
    Here, $\mathit{diff_{value}(P_{ku},C,K)}$ returns the \diffvalue for a given cluster $C$ and KU $K$ from the KU-based expertise profile dataset $P_{ku}$ (see Step 1). The $Subset(P_{ku},K)$ corresponds to the column in $P_{ku}$ that matches the given KU $K$. The $Subset(P_{ku},C,K)$ filters $Subset(P_{ku},K)$ by selecting only those rows that are associated with developers from the cluster $C$. 
	%$Subset(P_{ku},C,K)$ is thus a subset of $Subset(P_{ku},K)$, targeting the developers in cluster $C$.

    Next, we plot the \diffvalue of every KU of each cluster in the form of a bar plot. The interpretation of the bar plot is straightforward. A very narrow bar of a KU within a cluster indicates that the median value of the KU in the cluster is very close to the overall median of the KU. Conversely, a very wide bar of a KU within a cluster indicates that the median of the KU in the cluster is very different from the overall median. Hence, wide bars call for special attention. To determine if a bar is too wide, we take $median(Subset(P_{ku},C,K)$ and determine whether it is outside the interquartile range (Q1 to Q3) of $Subset(P_{ku},K)$. In the positive case, the bar is emphasized with a vivid (non-transparent) color. 

	\item \textit{Step 3) Study the dispersion of clusters' size.} We calculate the Gini index~\citep{gini_index} for the distribution of cluster sizes. The Gini index is a measure of statistical dispersion among values of a frequency distribution. This index ranges from 0.0 to 1.0, with 0.0 representing perfect equality and 1.0 representing perfect inequality. A Gini index close to one indicates greater inequality in the size of the clusters (e.g., a single cluster containing most of the observations).
\end{itemize}

\begin{figure}[!t]
	\centering
	\includegraphics[width=1\linewidth]{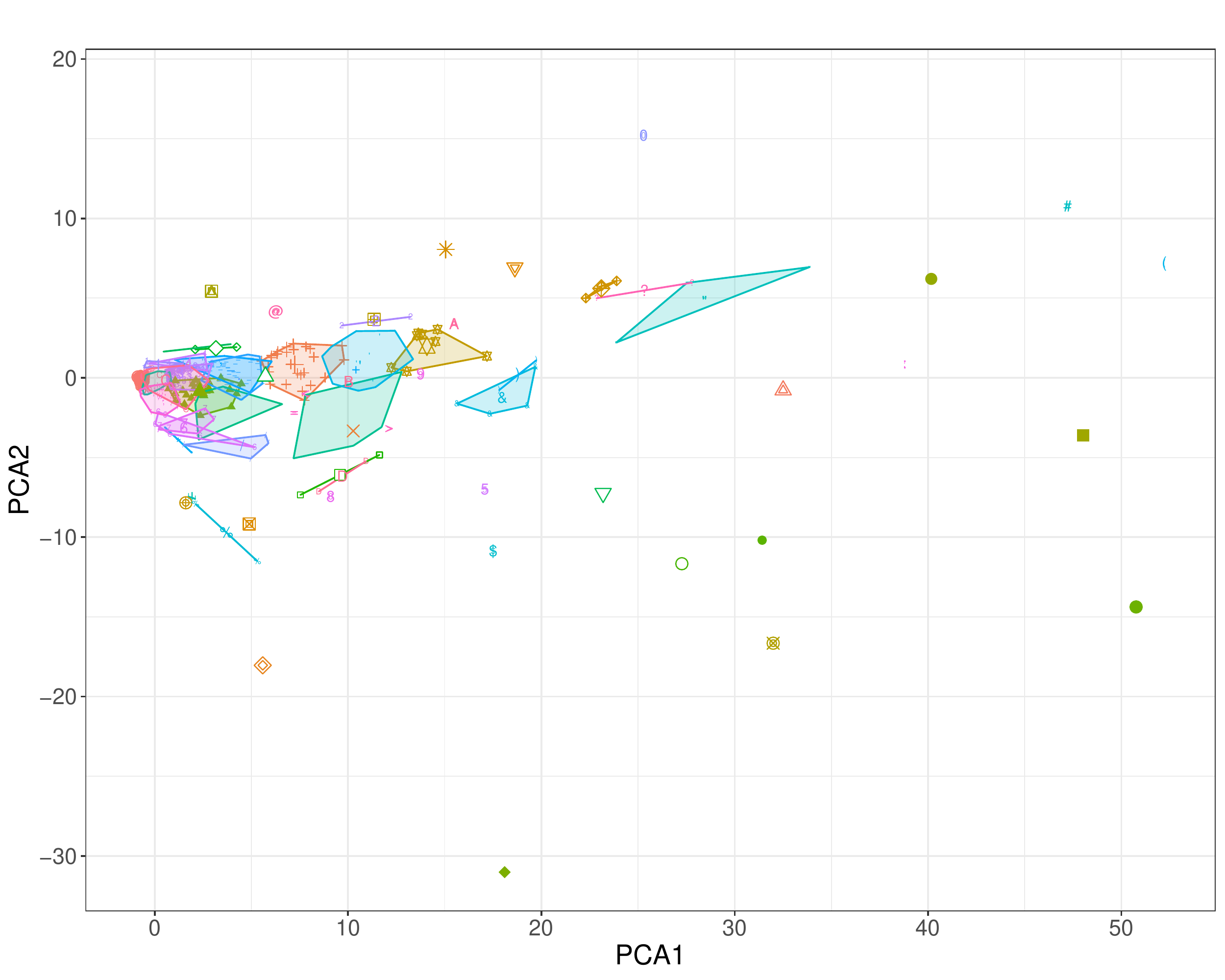}
	\caption{Visualization of developer clusters with KU-based expertise profiles. Polygons represent clusters and symbols represent developers.}
	\label{fig:prel_study-A-ku-clusters}
\end{figure}

\begin{figure*}[!t]
	\centering
	\includegraphics[width=1\linewidth]{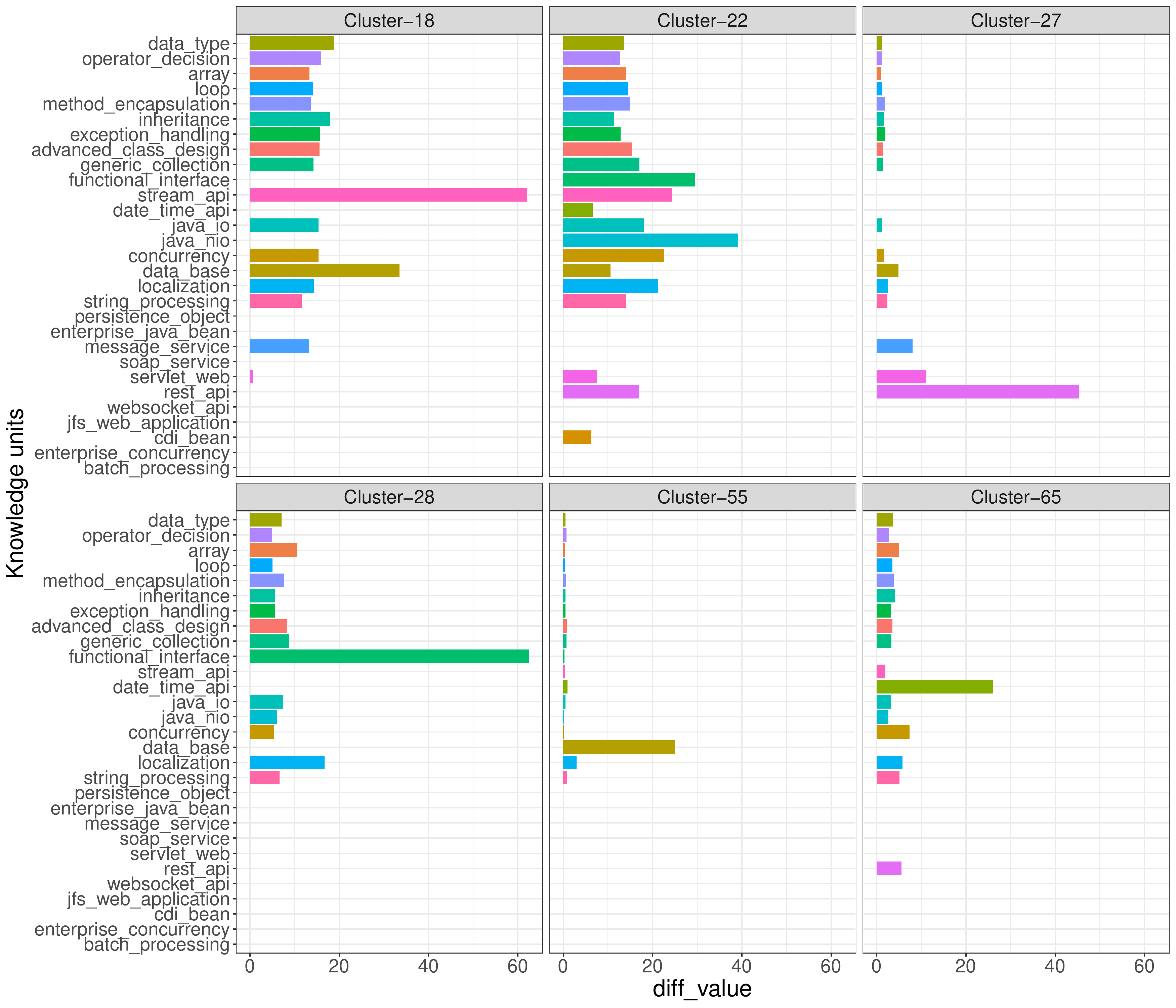}
	\caption{Visualization of the characteristics of the identified clusters via knowledge units.}
	\label{fig:prel_study-A-cluster-characteristics}
\end{figure*}

\smallskip \noindent \textbf{Findings.} \observation{KU-based expertise profiles form 71 different clusters of developers, indicating that KUs offer a fine-grained lens to study developers’ expertise.} Figure~\ref{fig:prel_study-A-ku-clusters} presents a visualization of the 71 developer clusters that are formed with KU-based expertise profiles. We observe that the generated clusters are separated from each other. We also observe that the number of cluster members are different across the clusters. The Gini index for clusters' size is 0.96, indicating that there exists very strong inequality (i.e., a few clusters contain most of the developers). Indeed, we observe that the largest cluster contains 80\% of all developers. That is, the majority of the studied developers have similar KU expertise. We also observe that 57\% (41 out of 71) of the generated clusters are singleton clusters (i.e., the size of these clusters is one). These singleton clusters are special because the single developer in them has a unique KU-based expertise profile compared to all other studied developers. 

%Interestingly, such imbalance is a recurrent phenomenon in several developer classification studies~\citep{goeminne2011evidence, joblin2017classifying}. 

%\goliva{do these guys belong to different projects?} \ahsan{yes, these guys are from 9 out of our 10 projects} \goliva{we can discuss this angle a bit more if reviewers are interested.}

% apache_activemq            7
%2 apache_groovy             5
%3 apache_hbase              2
%4 apache_hive               3
%5 apache_storm              8
%6 apache_wicket             4
%7 caskdata_cdap             6
%8 elastic_elasticsearch     2
%9 jruby_jruby               4

\smallskip \observation{Clusters generated from KU-based expertise profiles reveal developers with different code expertise.} We observe that each cluster has its own intrinsic characteristics derived from the incidence level of each KU. To illustrate our point, we randomly select six clusters and visualize their characteristics in Figure~\ref{fig:prel_study-A-cluster-characteristics}. The vivid colored KU bar in a cluster indicates that the median value of the KU in that cluster is outside the interquartile range (Q1 to Q3) of the median of the KU in the entire dataset. Hence, these KUs with wide bars indicate important characteristics of the cluster. The wider a vivid coloured KU bar is, the more prominent the KU is among the developers of that particular cluster. As we observe Figure~\ref{fig:prel_study-A-cluster-characteristics}, the ``Stream API KU'' is wider among the vivid colored KU bars in Cluster-18. As such, the ``Stream API KU'' is prominent among developers of Cluster-18 and an important characteristic. Similarly, ``Java NIO KU'' is prominent in Cluster-22, ``Rest API KU'' is prominent in Cluster-27, ``Functional Interface KU'' is prominent in Cluster-28, ``Database KU'' is prominent in Cluster-55 and ``Date Time API KU'' is prominent in Cluster-65. Such a prominent KU of a cluster indicates that developers of that cluster encounter the KU more frequently than other developers while performing code changes. Hence, developers of the cluster have more expertise in code changes related to that particular wide vivid coloured KU. These findings show that clusters vary from one another and each cluster reveal developers with different code expertise. 

\begin{footnotesize}
	\begin{mybox}{Summary}
		\textbf{\PrelimStudy}
		\tcblower
		Yes. Through the lens of KUs, we can identify 71 different clusters of developers. Each cluster hosts a set of developers with unique KU-based expertise profile.
	\end{mybox}
\end{footnotesize}
    \section{A study of code reviewer recommendation using KUs}
\label{sec:Main_study}

In the following, we address our three research questions. For each research question, we discuss our motivation for studying it, the approach used to answer it, and the findings that we observe.

\subsection{RQ1: \RQOne}
\label{sec:KU_Recommendation_System}

\noindent \textbf{Motivation.} As we discuss in Section~\ref{sec:Intro}, mapping different expertises to individual developers is a key requirement for effective code review assignment. As we show in the preliminary study, KUs can capture developers' programming language expertise. Hence, in this research question we build a KU-based code reviewer recommender system called KUREC and evaluate its performance. KUREC takes into the code contributions (commits) and prior code reviews performed by an individual when issuing reviewer recommendations. 

We clarify that our goal is \textit{not} to design the most accurate code recommender system possible, but rather to determine where its performance stands in comparison to solid baselines. Such a rationale derives from the novelty surrounding the conceptualization and operationalization of KUs.

\begin{figure*}[!t]
    \centering
    \includegraphics[width=0.9\linewidth]{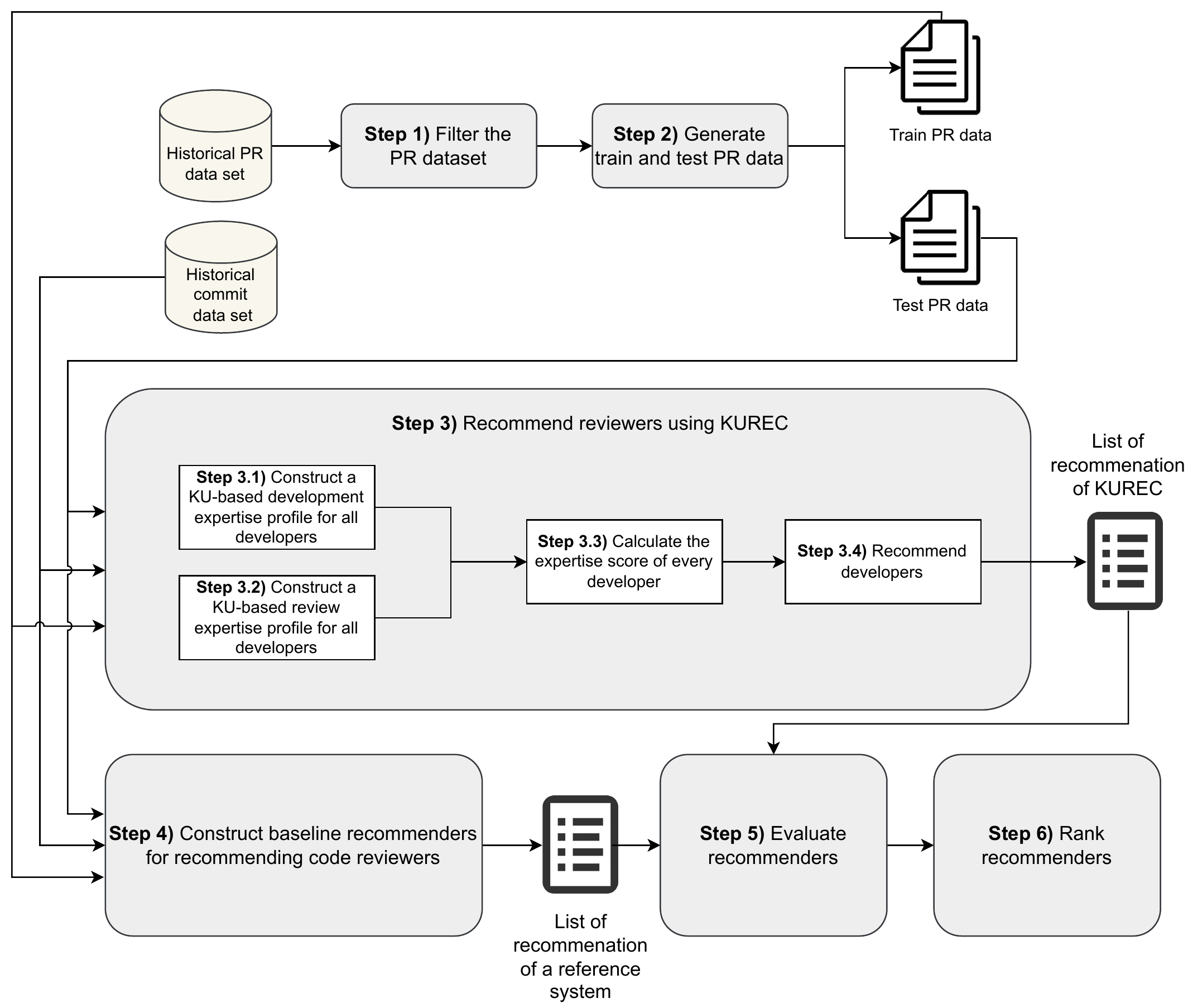}
    \caption{Our approach for building and evaluating KUREC.}
    \label{fig:rq3-recommendation-systems}
\end{figure*}

\smallskip \noindent \textbf{Approach.} We use pull requests to train and test both KUREC and other four baseline recommenders. \textbf{The key assumption of KUREC is that the code reviewers of a given PR are most likely experts in the KUs that appear in the changed files of the given PR.} Our approach for building and evaluating KUREC is summarized in Figure~\ref{fig:rq3-recommendation-systems}. In the following, we describe each step of our approach in detail.

% \goliva{in threats, you need to discuss that our approach assumes that reviewers will be developers and this might not always be the case in reality. still, our performance is super good and combining KUREC with other recommenders lead to even better results.}\ahsan{our approach considers both development expertise and review expertise. So, the confusion arises whether we should use the terminology `developers' vs `reviewers'. I believe this is not a threat. We need to carefully decide which terminology we should use `developers' vs `reviewers' or `contributor'. Do you have any suggestion?.}

\begin{itemize}[itemsep=3pt, wide = 0pt, topsep=1pt, listparindent=\parindent, label=$\bullet$]
    \item {\textit{Step 1) Filter the PR dataset.}} In section \ref{sec:Data_Collection}, we discussed how we collected pull request (PR) data. Since our objective is to build a recommender for \textit{code} reviewers, we only select those PRs that (i) include changed Java files, (ii) had at least one reviewer assigned to them, and (iii) are in a \texttt{closed} status. To ensure that we have enough PR data for training and testing a recommender, we only select those studied projects that contain at least 100 PRs. As a result of applying this filter, we remove the Cdap and Jruby projects from our study. As a result, we obtain a PR dataset for eight projects.

    \item {\textit{Step 2) Generate train and test PR data.}} We need to split the PR data for training and testing the recommenders. To split the PR dataset of a given project, we first sort the dataset chronologically. Our rationale is to avoid training recommenders with future data and testing those projects with past data (data leakage problem \citep{campos2011towards}). Then, in accordance with several prior studies~\citep{xia_recom_2017,yu_reviewer_IST_2016,ouni_search_ICSME_2016}, we split each dataset into a training dataset containing the initial 80\% of the data and a testing dataset containing the remaining 20\% of the data.
    
    %REMINDER: \goliva{shouldn't you be using a sliding window approach?} \ahsan{I think, sliding window can be important if we want to see how much historical data is required for the systems. I think this is a different research question. Since previous studies also follow this data splitting technique, I think we can defend present splitting approach. What do you think?}

	\item {\textit{Step 3) Recommend reviewers using KUREC.}} For each PR in the test dataset, we perform the following substeps to build KUREC and get its recommendations: 
    
    %\ahsan{everything is consistent except that we did not mention that we only used the KUs that are found in the changed file of a given test PR to calculate the KU expertise for recommending reviewers of the test PR. There was a first step that mentioned that we only extracted the KUs that are found in the changed files of a given PR.}
    
    \begin{itemize}[itemsep=3pt, wide = 0pt, topsep=0pt, listparindent=\parindent]
       
        \item \textit{Step 3.1) Construct a KU-based development expertise profile for all developers.} We first select all commits that are performed prior to the opening date of the PR in question. Then, we follow the same approach that we present in Step 3 of Section~\ref{sec:Data_Collection} to construct a KU-based development expertise profile for every developer. The collection of all profiles is represented with a matrix $DevExpKU(D_{i},KU_{j})$, where $D_{i}$ refers to a developer $i$ that performed at least one of the selected commits and $KU_{j}$ refers to one of the 28 KUs listed in Table~\ref{tab:topic_definition}.

        \item \textit{Step 3.2) Construct a KU-based review expertise profile for all developers.} We first select all those PRs that were created prior to the given PR. Then, we collect all changed files that are requested for review in each of the selected PRs. Next, we proceed analogously to Step 3.1, which results in the construction of a KU-based reviewer expertise profile for every developer. The collection of all these profiles is represented with a matrix $RevExpKU(D_{i},KU_{j})$, where $D_{i}$ refers to a developer $i$ that performed at least one code review for the selected PRs and $KU_{j}$ refers to one of the 28 KUs listed in Table~\ref{tab:topic_definition}.
        %every PR as code review: https://gitential.com/pull-requests-and-code-reviews/

        \item \textit{Step 3.3) Calculate the expertise score of every developer.} A developer who recently read or worked on a piece of code is likely to have fresher and clearer memories about the properties of such code (including the KUs that appear therein)~\citep{dec_knowledge_time}. It is thus not surprising that the recency of code changes and reviews has been taken into account by several reviewer recommenders~\citep{chrev_automatic_review_recommend_ASE_2015,ouni_search_ICSME_2016}. KUREC also takes recency into account to calculate the final expertise score of every developer.
        
        Following our terminology, let $DevExpKU(D{_i})$ be the KU-based development expertise profile of a developer $D_i$. KUREC adds a \textit{development recency bonus} to each $KU_{j}$ in $DevExpKU(D{_i})$ by (a) determining the last time that $D_i$ committed a file containing $KU_{j}$ and (b) judging how recent that is. More specifically, let $C$ be the last commit by ${D_{i}}$ that includes a file with $KU_{j}$. The recency bonus of $KU_{j}$ for $D_{i}$ is calculated as the \textit{inverse of the number of days between the commit date of $C$ and the opening date of the PR in question}. Formally, we represent the development recency bonus of $KU_{j}$ for developer $D_{i}$ as $RecDevKU(D_{i},KU_{j})$. 
        
        Analogously, KUREC also adds a \textit{review recency bonus} to each $KU_{j}$ in $RevExpKUs(D{_i})$. Let $R$ be the last code review by $D_{i}$ in which some reviewed file contained $KU_{j}$. In this case, the recency bonus of $KU_{j}$ for $D_{i}$ is calculated as the \textit{inverse of the number of days between the opening date of the PR that underwent $R$ and the opening date of the PR in question}. We represent the review recency bonus of $KU_{j}$ for developer $D_{i}$ as $RecRevKU(D_{i},KU_{j})$. Lastly, KUREC attributes a final KU-based expertise score $ExpertiseScore(D_{i})$ to each developer as follows:
        %\goliva{we should be using the review date instead. although a review might take some days to finish. so ideally, we would like to capture the review end date or something} 
        
        \vspace{-0.6cm}
        \begin {align*}
        & \scriptstyle DevScore(D_{i}) = \sum_{j = 1}^{N} [DevExpKU(D_{i}, KU_{j}) + RecDevKU(D_{i},KU_{j})]
        \end{align*}

        \vspace{-0.7cm}
        \begin {align*}
        & \scriptstyle RevScore(D_{i}) = \sum_{j = 1}^{N} [RevExpKU(D_{i}, KU_{j}) + RecRevKU(D_{i},KU_{j})]
        \end{align*}

        \vspace{-0.75cm}
        \begin{align*}
            \scriptstyle ExpertiseScore(D_{i}) = DevScore(D_{i}) + RevScore(D_{i})
        \end{align*}

        Here, $1 \leq N \leq 28$ is the number of KUs that appear in the changed files of the PR in question (i.e., a PR from the test dataset). Our rationale is that a reviewer of a given PR should be an expert in the KUs that appear in the changed files of such a PR.

        \item \textit{Step 3.4) Recommend developers.} We recommend the top \textit{K} ranked developers based on their expertise score, where \textit{K} ranges from 1 to 5.
    \end{itemize}

    \item {\textit{Step 4) Construct baseline recommenders for recommending code reviewers.}} To evaluate the performance of KUREC, we construct four different reviewer recommenders: (a) a commit-frequency-based recommender (CF)~\citep{dev_interaction_exp_MSR_2013}, (b) a review-frequency-based recommender (RF)~\citep{who_should_review_pull_request}, (c) a modification-expertise-based recommender (ER)~\citep{exp_recom_cscw, code_rev_recom_empirical_ASE_2016} and (d) a review-frequency-based recommender (CHREV)~\citep{chrev_automatic_review_recommend_ASE_2015}. We build the CF recommender because \textit{commit frequency} has frequently been used as an indicator for development expertise~\citep{dev_interaction_exp_MSR_2013}. We choose RF as a baseline for its simplicity to build the recommender~\citep{who_should_review_pull_request}. We select ER as other baseline recommender because it has been widely used in other studies~\citep{exp_recom_cscw, exp_browser_ICSE_2002, code_rev_recom_empirical_ASE_2016, eval_exp_recom_GROUP_2001}. Finally, we choose the CHREV recommender for two reasons. First, CHREV has been shown to outperform three other recommenders, namely REVFINDER~\citep{thongtanunam2015should}, xFinder~\citep{kagdi2008}, and REVCOM. Second, the steps to reimplement CHREV are clearly described and easy to follow~\citep{chrev_automatic_review_recommend_ASE_2015}. We briefly describe the construction process of these four baseline recommenders below.
    
    \noindent \item [(i)] Commit-frequency-based recommender (CF): For a given PR in the test dataset, CF first selects all the commits that are performed before the opening date of the given PR. Next, the recommender counts the number of commits that are performed by each developer. Finally, the recommender sorts developers in decreasing order of commit counts and recommends the top-k ones.
    
    \noindent \item [(ii)] Review-frequency-based recommender (RF): For a given PR in the test dataset, RF first selects all the PRs that are created prior to the opening date of the given PR. Then, RF counts the number of reviews per developer. Next, developers are sorted in decreasing order of review count. Finally, RF recommends the top-k ranked developers.

    \noindent \item [(iii)] Modification-expertise-based recommender (ER): The ER recommender relies on the assumption that a developer who \textit{last} modified a source code file is the most appropriate team member to review new changes to that same file~\citep{eval_exp_recom_GROUP_2001}. The rationale is that such a developer knows that file and potentially has fresh memories of its characteristics (e.g., structure, dependencies, behavior, and purpose). The ER recommender has been considered in several prior studies~\citep{exp_recom_cscw, exp_browser_ICSE_2002, code_rev_recom_empirical_ASE_2016, eval_exp_recom_GROUP_2001}. We adapt the ER recommender for our context as follows. For a given PR in the test dataset, ER first identifies all the changed files in the PR. We refer to this set of files as S. Next, a list of developers who modified any of those files in the past (i.e., in commits preceding the PR in question) is generated. Then, ER sorts developers in reverse chronological order based on the date of their last changed file in S. Finally, ER recommends the top-k ranked developers. 

    \noindent \item [(iv)] Review-history-based recommender (CHREV): For a given PR $p$ in the test dataset with a changed file $f$, CHREV tracks the contribution of each past reviewer of that file across all PRs preceding $p$. CHREV distills review contribution into three measures: (1) total number of review comments contributed to previous code changes in all PRs prior to the given one (C), (2) total number of workdays (a work day is considered as a day on which a reviewer contributed at least one review comment to at least one changed file) (W) and (3) recency of the review comments (T). To calculate these measures for a changed file $f$ of PR $p$, all PRs preceding $p$ and that contain the changed file $f$ are used. Finally, a contribution factor termed xFactor is calculated based on these three measures as follows:

    \vspace{-2ex}
    \begin {align*}
     {xFactor(r,f) =
    \begin{cases}
       \frac{C_f}{C'_f} + \frac{W_f}{W'_f} + \frac{1}{|T_f - T'_f|} \quad if\  |T_f - T'_f| > 0 \\ \\
       \frac{C_f}{C'_f} + \frac{W_f}{W'_f} + 1 \quad \quad \quad \ \ if \ |T_f - T'_f| = 0
    \end{cases}}
    \end{align*}

    Here, $C_{f}$ is the number of review comments that are contributed by a reviewer \textit{r} for the file \textit{f} and $C'_{f}$ is the total number of review comments that are written for the file \textit{f}. $W_{f}$ is the number of workdays of the reviewer r on which they contributed review comments for the file f and $W'_{f}$ is the total number of workdays on which review comments were contributed for the file f. $T_{f}$ is the most recent workday of the reviewer r with the file f and $T'_{f}$ is the most workday on which a review comment was contributed for the file f.
    
    %\goliva{for this paragraph, what I don't understand is whether you're considering only the PR in question or whether these things are being computed based on all PRs that contained the file f. Please clarify}\ahsan{These are being contributed based on all PRs that contained the file f. Added a sentence in the above paragraph to clarify this point. Fixed.} 
    
    Finally, the xFactor for all the changed files in the given PR are aggregated for each reviewer, and a score is generated for all the reviewers who contributed to the review of these changed files in past PRs prior to the given one. The final score is calculated as follows:

    \vspace{-2ex}
    \begin {align*}
    Score(r) = \sum_{i = 1}^{N} xFactor(r,f_{i})
    \end{align*}
    
    The reviewers are ranked in descending order of their final score values and the top-K reviewers are recommended. %\ahsan{rewrite the section to introduce how CHREV works. Please check if it is clear now.}

    \item {\textit{Step 5) Evaluate recommenders.}} To evaluate the performance of recommenders, we use top-k accuracy~\citep{xia2015should,thongtanunam2015should,chrev_automatic_review_recommend_ASE_2015}, and MAP~\citep{he2020diversified,devrec_TSE_2021}, as they are both widely used in the evaluation of recommenders within software engineering. We calculate these measures using the same formulations as~\citet{devrec_TSE_2021}):
    
    \vspace{-2ex}
    \begin{align}
        \textit{Top-k accuracy} = \frac{\sum_{r\in R}^{}isCorrect(r,Top-k)}{|R|}
    \end{align}
    
    Here, $R$ denotes the set of PRs in the test dataset. The $isCorrect(r,Top-k)$ returns 1 if at least one of top-k developers is the correct reviewer of the PR $r$ and returns 0 otherwise.
    
    %\goliva{I don't understand exactly what this means --> The top-k accuracy cannot measure how good the recommender's performance is in ranking the recommended reviewers. make it clearer and/or provide an example}\ahsan{top-k accuracy only check how many of the test PRs, the recommender correctly recommends reviewers. However, top-k accuracy can not measure if the ranking is good/bad. For example, the correct reviewer of a PR is $r_{a}$ and two recommenders (A and B) recommend top-3 reviewers. Say recommender A recommends $<r_{d},r_{e},r_{a}>$ and recommender B recommends $<r_{a},r_{b},r_{d}>$. The top-k accuracy can not distinguish which one is better. Of course recommender B's ranking is better because it can recommend the true reviewers at the top two ranked position. Whereas recommender A recommends one true reviewer at the third position. I updated the highlighted line to make it clear. FIXED} 
    
    The top-k accuracy cannot measure how good the recommender's performance is in ranking the recommended reviewers. For example, the correct reviewer of a PR is $r_{a}$ and two recommenders (A and B) recommend top-3 reviewers. Now, A recommends $\langle r_{d},r_{e},r_{a}\rangle$ and B recommends $\langle r_{a},r_{b},r_{d} \rangle$. The top-k accuracy is equal for both A and B recommenders because they correctly recommend one reviewer ($r_{a}$). However, in terms of ranking the correct reviewer, recommender B's ranking is better than A's ranking. Because recommender B recommends the correct reviewers at the top one ranked position. Whereas recommender A recommends the correct reviewer at the third position. To understand how good the recommender's performance is in ranking the recommended reviewers, we calculate the Mean Average Precision (MAP) metric~\citep{he2020diversified,devrec_TSE_2021}. MAP assigns higher importance to the results at top ranks. To calculate MAP, we first calculate the Average Precision (AP):

    \vspace{-2ex} 
    \begin{align}
        AP@k = \frac{\sum_{i = 1}^{k}\frac{s(i)}{i}\times rel(i)}{\sum_{i = 1}^{k}rel(i)}
    \end{align}

    Here, $i$ is the position of each developer in the recommended list of developers, and $s(i)$ is the sequence number of the correct developer at position $i$. The $rel(i)$ returns 1 if the $i_{th}$ developer in the list is correct and 0 otherwise. $MAP@k$ is the average of $AP@k$ over all the PRs in the test dataset. For example, we recommend $K=5$ developers (d1, d2, d3, d4, d5) and the 1-st, 3-rd and 5-th developers are correct. The sequence number of d1, d3, and d5 are 1, 2, and 3, i.e., s(1) = 1, s(3) = 2, and s(5) = 3. Then, the AP@5 can be calculated as $(\frac{1}{1} +\frac{2}{3}+\frac{3}{5})$. MAP@5 averages the AP@5s across all the recommended lists of the PRs in the test dataset.
    
    \item {\textit{Step 6) Rank recommenders.}} To rank the recommenders for a given evaluation metric (e.g., MAP), we first group the results of a recommender across all top-k recommendations (for \textit{k} in 1 to 5) and studied projects. This grouping leads to a distribution with 40 data points (5 recommendations $\times$ 8 studied projects). Next, we apply the Scott-Knott ESD technique (SK-ESD)~\citep{ghotra_feature_importance_ICSME, mittas_ranking_feature_TSE, kla_model_validation} to obtain a ranking of the recommenders. The SK-ESD is a method of multiple comparison that leverages a hierarchical clustering to partition the set of treatment values (e.g., medians or means) into statistically distinct groups with non-negligible difference~\citep{mittas_ranking_feature_TSE}.
    
    %\goliva{need to say a few words about what SK is and what it does. copy from paper 1. A reviewer in paper 1 also had a comments about SK (about the non-parametric aspect I think). so please tackle it here as well.)}\ahsan{We did not mention details of the SK-ESD in paper 1. Added one sentence about what SK-ESD is.} We use the SK-ESD implementation provided by Tim Menzies, which employs non-parametric statistical testing (i.e., Mann-Whitney U-test) and Cliff's Delta~\citep{cliff_delta} as a measure of effect size~\citep{tim_sk}\ahsan{The reviewer concerns was about which non-parametric test was used in Tim's implementation. not sure if Tim uses Mann-Whitney U-test?Could you please confirm it?.}.
\end{itemize}

\begin{figure}[!ht]
    \centering
    \includegraphics[width=1\linewidth]{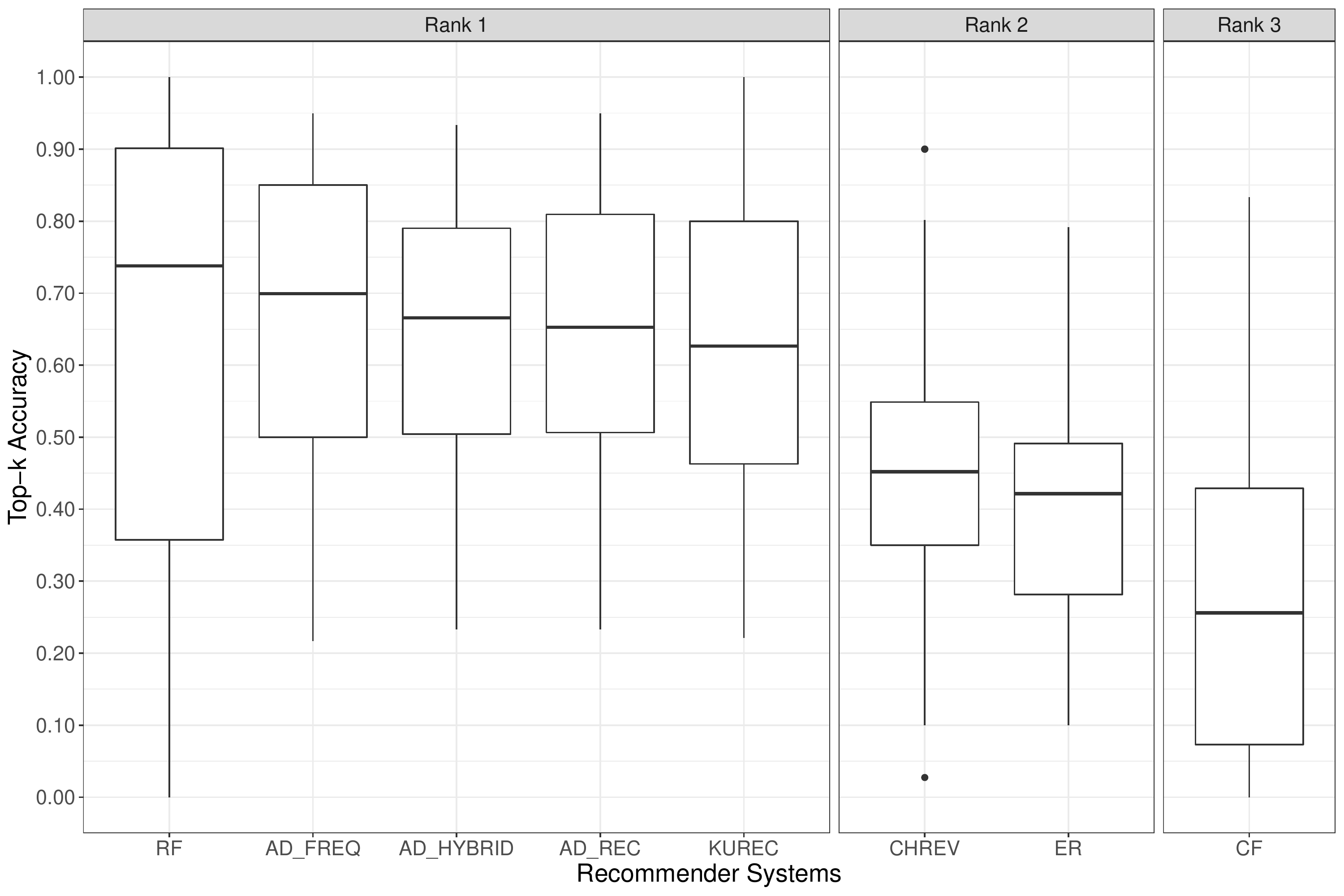}
    \caption{The distribution of top-k accuracy (k changes from 1 to 5) of the recommenders across the studied projects. Recommender systems are grouped according to their Scott-Knott ESD rank.}
    \label{fig:rq3-sk-rank-accuracy}
\end{figure}

\begin{figure}[!ht]
    \centering
    \includegraphics[width=1\linewidth]{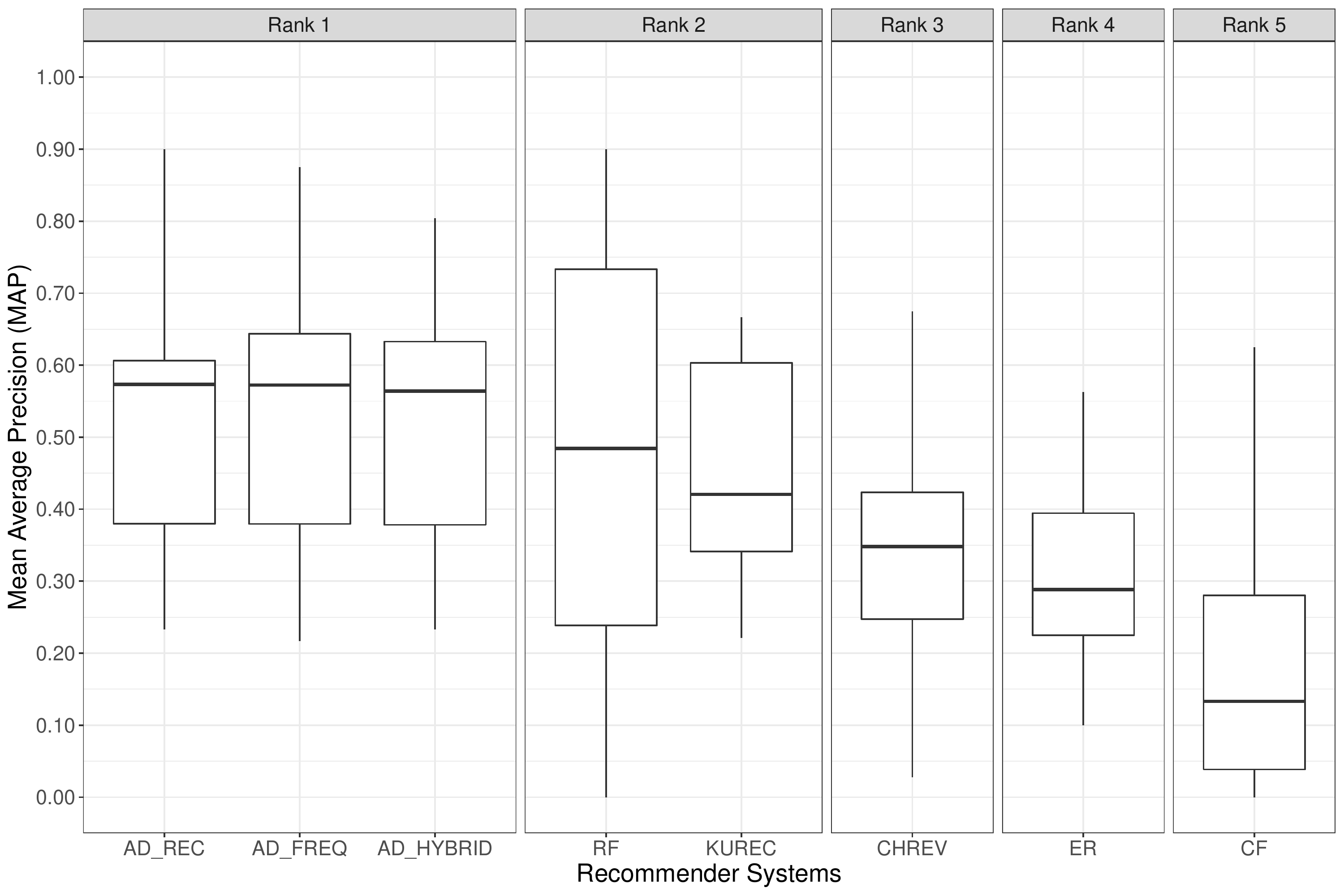}
    \caption{The distribution of MAP@k (k changes from 1 to 5) of the recommenders across the studied projects. Recommenders are grouped according to their Scott-Knott ESD rank.}
    \label{fig:rq3-sk-rank-map}
\end{figure}

\smallskip \noindent \textbf{Findings.} \observation{KUREC performs as well as the top-performing baseline recommender (RF) and outperforms the remaining three baselines.} Figure~\ref{fig:rq3-sk-rank-accuracy} presents the distribution of top-k accuracy and Figure~\ref{fig:rq3-sk-rank-map} presents the distribution of MAP across the studied projects. The recommenders are grouped according to their Scott-Knott ESD rank. We observe that our KUREC achieves the highest rank along with the RF among the baseline recommenders in both top-k accuracy and MAP. We observe that the absolute difference of the median top-k accuracy between the KUREC and the RF is only 0.08 and the absolute difference of median MAP between these two recommenders is only 0.05. Even though the medians differ a little, the performance of our KUREC and the RF, the top-performing baseline recommender, is identical from an effect size standpoint. We also observe that our KUREC achieves a higher rank than the remaining three baseline recommenders (e.g., CHREV, ER and CF) in both top-k accuracy and MAP. Such a result indicates that our KUREC outperform these baseline recommenders and can be used effectively for recommending code reviewers in PRs.

\observation{The performance of our KUREC is more stable than that of the top-performing baseline recommender (RF).} As we observe from Figure~\ref{fig:rq3-sk-rank-accuracy} and Figure~\ref{fig:rq3-sk-rank-map}, the interquartile range of our KUREC is smaller than the RF in both top-k accuracy and MAP. For MAP, the interquartile range of our KUREC is only 0.25, whereas the interquartile range of the RF is 0.56. Such a result indicates that KUREC is more stable than the RF. 

We can also notice from Table~\ref{tab:rq3-accuracy} and Table~\ref{tab:rq3-map} that the performance of our KUREC is stable. Here, Table~\ref{tab:rq3-accuracy} presents the top-k accuracy results and Table~\ref{tab:rq3-map} presents the MAP results across the studied projects. We observe that the minimum top-k accuracy and MAP of our KUREC are always above 0.20, whereas these values could reach 0.00 for the baseline recommenders (see the last column of Table~\ref{tab:rq3-accuracy} and Table~\ref{tab:rq3-map}). For example, in Apache ActiveMQ, the minimum top-k accuracy and the MAP of CF and CHREV are 0.00. We can even notice unstable performance in the top-performing recommender (RF). The minimum top-k accuracy and MAP are 0.00 for the RF in Apache Storm and Elastic Search projects. In Elastic Search, the top-k accuracy of the RF is the worst among the baseline recommenders. On the other hand, our KUREC outperforms all the baseline recommenders including the RF. Such a result indicates that our KUREC is more stable than the RF. In any recommender, stability is a desired property because a stable recommender can gain users' trust and acceptance of recommendations~\citep{stability_recommendation_system}. Based on our aforementioned results, we can conclude that our KUREC is he best recommender among the baseline recommenders since our KUREC performs identically to the top one but is more stable.

% Table in seperate Latex: https://tex.stackexchange.com/questions/150829/save-table-on-a-separate-file

%%%%%%% [Accuracy Results]  %%%%%%%%
\begin{table}[]
    \scriptsize
    \caption{The top-k accuracy values of the studied recommenders.}
    \label{tab:rq3-accuracy}
    \resizebox{0.95\textwidth}{!}{
}  & \multirow{4}{*}{Baseline}                                                                               & CF                                                                                                   & 0.16                             & 0.25                             & 0.34                             & 0.42                             & 0.47                             & 0.16                                                                                         \\
                                                                                        &                                                                                                         & RF                                                                                                   & 0.03                             & 0.12                             & 0.17                             & 0.23                             & 0.29                             & 0.03                                                                                         \\
                                                                                        &                                                                                                         & ER                                                                                                   & 0.28                             & 0.42                             & 0.50                             & 0.51                             & 0.54                             & 0.28                                                                                         \\
                                                                                        &                                                                                                         & CHREV                                                                                                & 0.25                             & 0.42                             & 0.46                             & 0.52                             & 0.52                             & 0.25                                                                                         \\ \cmidrule{2-9}
                                                                                        & KU                                                                                                      & KUREC                                                                                                & 0.30                             & 0.43                             & 0.47                             & 0.53                             & 0.56                             & 0.27                                                                                         \\ \cmidrule{2-9}
                                                                                        & \multirow{3}{*}{Combined}                                                                               & AD\_FREQ                                                                                             & 0.28                             & 0.42                             & 0.50                             & 0.56                             & 0.59                             & 0.28                                                                                         \\
                                                                                        &                                                                                                         & AD\_REC                                                                                              & 0.27                             & 0.42                             & 0.51                             & 0.57                             & 0.60                             & 0.27                                                                                         \\
                                                                                        &                                                                                                         & AD\_HYBRID                                                                                           & 0.27                             & 0.42                             & 0.51                             & 0.56                             & 0.60                             & 0.27                                                                                         \\ \midrule \midrule
            \multirow{8}{*}{Median}                                                     & \multirow{4}{*}{Baseline}                                                                               & CF                                                                                                   & 0.11                             & 0.15                             & 0.19                             & 0.32                             & 0.36                             & 0.11                                                                                         \\
                                                                                        &                                                                                                         & RF                                                                                                   & 0.38                             & 0.64                             & 0.76                             & 0.85                             & 0.88                             & 0.38                                                                                         \\
                                                                                        &                                                                                                         & ER                                                                                                   & 0.24                             & 0.36                             & 0.42                             & 0.46                             & 0.49                             & 0.24                                                                                         \\
                                                                                        &                                                                                                         & CHREV                                                                                                & 0.29                             & 0.39                             & 0.46                             & 0.53                             & 0.56                             & 0.29                                                                                         \\ \cmidrule{2-9}
                                                                                        & KU                                                                                                      & KUREC                                                                                                & 0.29                             & 0.59                             & 0.69                             & 0.76                             & 0.84                             & 0.29                                                                                         \\ \cmidrule{2-9}
                                                                                        & \multirow{3}{*}{Combined}                                                                               & AD\_FREQ                                                                                             & 0.46                             & 0.64                             & 0.74                             & 0.78                             & 0.83                             & 0.46                                                                                         \\
                                                                                        &                                                                                                         & AD\_REC                                                                                              & 0.46                             & 0.61                             & 0.68                             & 0.71                             & 0.75                             & 0.46                                                                                         \\
                                                                                        &                                                                                                         & AD\_HYBRID                                                                                           & 0.43                             & 0.63                             & 0.70                             & 0.73                             & 0.77                             & 0.43                                                                                         \\ \bottomrule
            \end{tabular}
    }
    \end{table}
%%%%%%% [MAP Results]  %%%%%%%%
\begin{table}[]
    \scriptsize
    \caption{The mean average precision (MAP) values of the studied recommenders.}
    \label{tab:rq3-map}
    \resizebox{0.95\textwidth}{!}{
}  & \multirow{4}{*}{Baseline}                                                                               & CF                                                                                                   & 0.16         & 0.21         & 0.23         & 0.25         & 0.26         & 0.16                                                                                         \\
                                                                                        &                                                                                                         & RF                                                                                                   & 0.03         & 0.07         & 0.09         & 0.10         & 0.11         & 0.03                                                                                         \\
                                                                                        &                                                                                                         & ER                                                                                                   & 0.28         & 0.31         & 0.32         & 0.38         & 0.39         & 0.28                                                                                         \\
                                                                                        &                                                                                                         & CHREV                                                                                                & 0.25         & 0.30         & 0.32         & 0.32         & 0.33         & 0.25                                                                                         \\ \cmidrule{2-9}
                                                                                         & KU                                                                                                      & KUREC                                                                                                & 0.30         & 0.34         & 0.36         & 0.38         & 0.43         & 0.30                                                                                         \\ \cmidrule{2-9}
                                                                                        & \multirow{3}{*}{Combined}                                                                               & AD\_FREQ                                                                                             & 0.28         & 0.35         & 0.37         & 0.38         & 0.42         & 0.28                                                                                         \\
                                                                                        &                                                                                                         & AD\_REC                                                                                              & 0.27         & 0.35         & 0.37         & 0.38         & 0.38         & 0.27                                                                                         \\
                                                                                        &                                                                                                         & AD\_HYBRID                                                                                           & 0.27         & 0.34         & 0.37         & 0.38         & 0.38         & 0.27                                                                                         \\ \midrule \midrule
            \multirow{8}{*}{Median}                                                     & \multirow{4}{*}{Baseline}                                                                               & CF                                                                                                   & 0.11         & 0.13         & 0.14         & 0.15         & 0.16         & 0.11                                                                                         \\
                                                                                        &                                                                                                         & RF                                                                                                   & 0.38         & 0.52         & 0.53         & 0.53         & 0.53         & 0.37                                                                                         \\
                                                                                        &                                                                                                         & ER                                                                                                   & 0.24         & 0.30         & 0.32         & 0.32         & 0.32         & 0.24                                                                                         \\
                                                                                        &                                                                                                         & CHREV                                                                                                & 0.29         & 0.34         & 0.36         & 0.36         & 0.37         & 0.29                                                                                         \\ \cmidrule{2-9}
                                                                                        & KU                                                                                                      & KUREC                                                                                                & 0.29         & 0.46         & 0.50         & 0.51         & 0.51         & 0.29                                                                                         \\ \cmidrule{2-9}
                                                                                        & \multirow{3}{*}{Combined}                                                                               & AD\_FREQ                                                                                             & 0.46         & 0.56         & 0.58         & 0.60         & 0.60         & 0.46                                                                                         \\
                                                                                        &                                                                                                         & AD\_REC                                                                                              & 0.46         & 0.56         & 0.59         & 0.59         & 0.59         & 0.46                                                                                         \\
                                                                                        &                                                                                                         & AD\_HYBRID                                                                                           & 0.43         & 0.55         & 0.57         & 0.58         & 0.58         & 0.43                                                                                         \\ \bottomrule
            \end{tabular}
    }
    \end{table}

\begin{footnotesize}
    \begin{mybox}{Summary}
    	\textbf{RQ1: \RQOne}
        \tcblower
    	Yes. Our KUREC recommender is the best option among the baselines. In particular,
    	\begin{itemize}[itemsep = 3pt, label=\textbullet, wide = 0pt]
            \item KUREC performs as well as the top-performing baseline recommender (RF).
            \item KUREC has a more stable performance than the top-performing recommender (RF).
    	\end{itemize}
    \end{mybox}
\end{footnotesize}

\subsection{RQ2: \RQTwo}

\noindent \textbf{Motivation.} In Section~\ref{sec:KU_Recommendation_System}, we observe that each recommender has its own strength (e.g., the KUREC leverages both development expertise and review expertise)
%(e.g., the RF performs well when a few reviewers perform most of the reviews) 
to perform a better recommendation and its own weakness (e.g., the RF performs the worst when many reviewers perform the reviews in a large project) that results in bad recommendations. Therefore, in this section, we want to see if straightforward combinations of the KUREC with the baseline recommenders result in a better performing recommender.

\smallskip \noindent \textbf{Approach.} To construct a combined recommender by leveraging the recommendations of different recommenders, we are motivated by the work of \citet{malik2008supporting}. In this work, the authors propose adaptive meta heuristics that combine various previously researched heuristics and improve the performance of change propagation task (i.e.,  changes required to other entities of the software recommender to avoid bugs). %Similarly to their approach, we also constructed three adaptive meta heuristics that combine all the previously studied recommenders (see in Section~\ref{sec:KU_Recommendation_System}). 
We follow the same approach of~\citet{malik2008supporting}.  In this approach, all the recommenders uses a Best Recommender System Table (BRST) to track the best-performing recommender. 

We briefly explain each of the steps that we follow in this approach. First, all the studied recommenders are trained with the training dataset. Then, a recommender is randomly selected  to predict the reviewers of the first PR from the test data. Next, for testing each of the remaining PRs of the test dataset, we select a recommender from the BRST. Once a PR is completed for recommendation, we identify the best-performing recommender and update the BRST. To identify the best-performing recommender for a given PR, we generate a score for each of the studied recommenders by combining both accuracy and MAP. We run every recommender for all previous PRs from our test dataset and calculate the top-k accuracy and MAP. Then, we calculate the average of the accuracy and the average of the MAP for all k recommendations. Next, we average the accuracy and the MAP, and generate a single score for each of the studied recommenders to rank them. We identify the best-performing recommender that has the highest score. Finally, we need to update the BRST when a PR is completed for recommendation. We implement the following three techniques to update the BRST:

\begin{enumerate}[wide = 0pt, itemsep = 3pt, label=\textbf{\arabic*})]
    \item \textbf{Adaptive Frequency Technique (AD\textunderscore FREQ).} In this technique, the BRST is a frequency table that stores the count of how many times each recommender becomes the best-performing recommender among all the previous PRs of the test dataset. To update the BRST for a given PR, we increment the count by one to the best-performing recommender. In order to recommend reviewers for a new PR, we select the recommender that has the highest count in the BRST.
    \item \textbf{Adaptive Recency Technique (AD\textunderscore REC).} In this technique, the BRST only stores the best-performing recommender that is identified in the last PR. To update the BRST for a given PR, we replace the currently stored recommender to the identified best-performing recommender for the given PR. In order to recommend reviewers for a new PR, we select the recommender that is stored in the BRST.
    \item \textbf{Adaptive Hybrid Technique (AD\textunderscore HYBRID).} This technique leverages both the frequency and recency techniques. In this technique, the BRST is a frequency table that stores the count of how many times each recommender becomes the best-performing recommender among the last 10 previous PRs of the test dataset. We increment the count by one to the best-performing recommender for the given PR and update the BRST.  To recommend reviewers for a new PR, we select the recommender that has the highest count in the BRST among the last 10 previous PRs.
\end{enumerate}

\smallskip \noindent \textbf{Findings.} 
\observation{All the combined recommenders outperform individual recommenders.} We observe that all three combined recommenders perform the same from an effect size perspective, ranking first in both top-k accuracy (Figure~\ref{fig:rq3-sk-rank-accuracy}) and MAP (Figure~\ref{fig:rq3-sk-rank-map}). For top-k accuracy, the RF, the KUREC and the combined recommenders are all tied (i.e., they all rank first). However, similar to the KUREC, the interquartile range of the combined recommenders is smaller than that of the RF (see Figure~\ref{fig:rq3-sk-rank-accuracy}). Such a result indicates that the combined recommenders are more stable than the RF. 
For MAP, the performance of the combined recommenders is even better. As we observe from Figure~\ref{fig:rq3-sk-rank-map}, the combined recommenders are the only ranked first recommenders outperforming all individual recommenders. The median MAP of the combined recommenders are higher than all baseline recommenders including the RF and the KUREC (see also the last row of Table~\ref{tab:rq3-map}). For example, the AD$\_$FREQ has a 9\% improvement at top-1 recommendation and 7\% improvement top-5 recommendation from the highest performing baseline recommender (i.e., the RF recommender). In addition, the interquartile range of all three combined recommenders is smaller than the RF and is similar to the KUREC. As we observe from the last column of Table~\ref{tab:rq3-map}, the minimum MAP of the combined recommenders never goes below 0.22 (whereas the minimum MAP of the RF reaches 0.0). These results indicate that combining the KUREC with baseline recommenders in a straightforward manner outperform individual recommenders.

\begin{footnotesize}
    \begin{mybox}{Summary}
    	\textbf{RQ2: \RQTwo}
        \tcblower
    	Yes. Combining the KU-based recommender (KUREC) with the baselines in a straightforward manner results in better performing recommenders. In particular,
    	\begin{itemize}[itemsep = 3pt, label=\textbullet, wide = 0pt]
            \item All the combined recommenders ranked first among all individual recommenders in terms of both top-k accuracy and MAP.
            \item The combined recommenders outperform all individual recommenders in terms of MAP.
            \item The performance of all combined recommenders is more stable than that of the top-performing baseline recommender (RF)
    	\end{itemize}
    \end{mybox}
\end{footnotesize}
\subsection{RQ3: \RQThree}

%\ahmed{By definition of your metric.. any recency based recommender would just perform quite well since it is how you defined reasonable.. or am I missing something here?}\ahsan{I do not think so. Our rationale to pick last six months as a threshold is to justify that our approach of identifying developers for reviewing files correct (they have recent knowledge about updating the files). If we do not use the threshold than reviewers could ask question that these developers are not qualified for reviewing the files as their commits are very old. What do you think? }

\noindent \textbf{Motivation.} To measure the performance of KUREC and the other baseline recommenders, we needed a ground truth. To build our ground truth, we examined the review history of a project. We considered a recommendation to be correct when such a recommendation matched the individual who actually reviewed that PR. This is a standard practice in the code review literature~\citep{chrev_automatic_review_recommend_ASE_2015,thongtanunam2015should,kagdi2008,devrec_TSE_2021}. However, as we observe in RQ1, there are certain projects in which few developers review most of the PRs (c.f., top-5 accuracy/MAP of RF). We thus wonder whether the recommendations issued by our studied recommenders would have been \textit{reasonable} choices in practice when those differ from what happened in reality. Investigating this hypothesis sheds light into whether KUREC's recommendations would be a valid alternative to recommending the same developers over and over (e.g., to rebalance review effort across the team).

\smallskip \noindent \textbf{Approach.} We consider a recommendation to be \textit{reasonable} if the recommended individual had recent development experience with the majority of the files included in the PR in question. That is, if someone recently contributed changes to the majority of the files in a PR, we assume that that person is likely knowledgeable about those files and would thus be a reasonable reviewer for that PR. More specifically, our approach works as follows. 

First, we pick a recommender \textit{R}. Next, we select the PRs from the test dataset where \textit{R}'s top-1 recommendation \textit{D} does \textit{not} match the ground truth (the actual reviewer). Let \textit{P} be one of those PRs. For each \textit{P}, we collect all files from commits and PRs that were submitted by \textit{D} in the six months preceding the submission of \textit{P}. Let \textit{F} be this collection of files. Finally, if at least 50\% of the changed files in \textit{P} are included in \textit{F}, then we consider \textit{D} to be a reasonable recommendation. We choose six months as a time frame to ensure that the reviewer has very recent expertise with the files included in the PR in question. We report the percentage of \textit{reasonable} recommendations for KUREC and all the other studied recommenders (i.e., KUREC's variations and baseline recommenders).

\smallskip \noindent \textbf{Findings.} \observation{KUREC is the recommender with the highest percentage of reasonable recommendations (63.4\%).} The percentage of reasonable recommendations per recommender (for PRs in which the top-1 recommendation does not match the ground truth) is as follows: KUREC: 63.4\%, ER: 60.9\%, AD\_FREQ: 59.4\%, AD\_HYBRID: 54.3\%, AD\_REC: 54.2\%, CHREV: 32.7\%, CF: 25.4\%, and RF: 15.2\%. Interestingly, both ER and the combined recommender also have a high percentage of reasonable recommendations. However, the combined recommenders are considerably more aligned to the ground truth compared to ER (c.f., Figure~\ref{fig:rq3-sk-rank-accuracy} and~\ref{fig:rq3-sk-rank-map}). The AD\_FREQ combined recommender, in particular, strikes the best balance between sticking to the ground truth (best recommender from RQ2) and issuing reasonable recommendations when those deviate from that ground truth (third best recommender in this RQ). Overall, the aforementioned results provide positive empirical evidence to support the idea that either AD\_FREQ or KUREC's recommendations could be a valid alternative to recommending the same developers over and over. However, more in-depth studies would still be necessary to determine how acceptable those recommendations would be in practice.

To conclude this RQ, we make a final observation. One might notice that the very approach of this RQ could have been converted into a new recommender that would likely outperform all of our studied ones (including KUREC and AD\_FREQ) in terms of the ratio of reasonable recommendations. However, we clarify that such a recommender would have limited applicability in practice because it would \textit{operate exclusively on contribution recency}. In other words, it would only recommend the same individuals over and over. Such a characteristic is undesirable, since (i) it does \textit{not} promote knowledge and expertise sharing (e.g., it aggravates the consequences of developer turnover ~\citep{mirsaeedi2020mitigating}) and (ii) it prevents adequate review workload distribution. Modern recommenders, such as KUREC and CHREV account for other factors \textit{in addition to recency}. For instance, KUREC considers KU-based development expertise when recommending reviewers. As a consequence, KUREC is able to recommend an individual even if that individual has never performed a code review before.

%Let's keep this here in case a reviewer wants to see an example
%\smallskip \noindent \textit{Illustrative example}: As an illustrative example, we discuss a pull request from ElasticSearch (PR\#77464)\footnote{\url{https://github.com/elastic/elasticsearch/pull/77464}} that fixes a date-related issue. To fix the issue, a single file was modified: ``CanMatchPreFilterSearchPhaseTests.java''. We observe that KUREC's recommendation (Jim Ferenczi\footnote{\url{https://github.com/jimczi?tab=overview\&from=2021-12-01\&to=2021-12-31}}) does not match the actual reviewer for that PR. Yet, we note that Jim has experience with that file. Jim not only had modified that same file in a prior commit\footnote{\url{https://github.com/elastic/elasticsearch/commit/2c7039}}, but also reviewed it in some other prior PR\footnote{\url{https://github.com/elastic/elasticsearch/pull/65583/files\#r542380492}}. Such an example illustrates that KUREC's recommended reviewers are indeed capable of reviewing PRs. 

\begin{footnotesize}
    \begin{mybox}{Summary}
    	\textbf{RQ3: \RQThree}
        \tcblower
    	KUREC is the recommender with the highest percentage of reasonable recommendations (63.4\%). Yet, AD\_FREQ strikes the best balance between sticking to the ground truth (c.f., RQ2) and issuing reasonable recommendations when those deviate from that ground truth (59.4\% reasonable recommendations).
    \end{mybox}
\end{footnotesize}
	%\ahmed{I wonder if we should move this related work to the fron and call it ``related work and background''}\ahsan{Fixed. Move it after the Introduction. Not sure if we need to add any Background of Reviewe mechanism. Just rename it to ``Related Work and Background'' as Ahmed mentioned. What do you think?} \goliva{Undid the related work move. There is already too much material before the RQs. Also, related work refers to paper results and diagrams, so putting it early on without doing modifications to the text doesn't work.}

\section{Related Work}
\label{sec:Related_Work}

\noindent \textbf{Reviewer recommendation.} \citet{chrev_automatic_review_recommend_ASE_2015} develop an approach, namely cHRev, that can automatically recommend reviewers based on their historical review contributions. The cHRev first extracts each source code file and generates a reviewer's expertise factor based on their historical contributions. Then the cHRev recommends reviewers based on the score of the cumulative expertise factors. The authors evaluate the performance of cHRev on three open-source systems and a commercial codebase at Microsoft. The results show that cHRev outperforms the existing state-of-the-art works for peer reviewer recommendations.

\citet{yu2016reviewer} investigate the effectiveness of traditional approaches to bug triaging and code review for pull-request reviewer recommendation. The authors also propose a recommendation model for reviewers of pull requests by analyzing comment networks. The results show that the performance of traditional approaches can be improved by combining them with the information extracted from prior social interactions (e.g., comments) between developers.

\citet{rahman2016correct} propose a code reviewer recommendation tool that recommends reviewers for a pull request by analyzing a developer's past experience with external software libraries and specialized technology (e.g., Google App Engine) that are used by the pull request. The authors evaluate the tool using 13,081 pull requests from 10 subject systems. The results show that the performance of the tool is promising, according to relevant literature.

\citet{thongtanunam2015should} propose a reviewer recommendation approach, namely REVFINDER, which is based on the past reviews of files with similar names and paths. The authors evaluate their performance on a case study of 42,045 reviews of four popular projects. The results show that REVFINDER achieves an accuracy of 79\% at top 10 recommendations and the overall ranking of REVFINDER is three times better than a baseline approach.

\citet{xia2015should} propose another approach for reviewer recommendation, namely TIE. TIE is based on the intuition that the same reviewers are likely to review code changes containing similar terms (i.e., words) and are like to review the same files or the files in a similar location. The evaluation results show that TIE outperforms the REVFINDER on open source systems.

Our work is different from the above studies because we introduce KUs and develop a recommender system using KUs that shows effectiveness in recommending code reviewers for PRs.

%[Excerpt that used to be in the introduction] A recent work by \citet{expert_represent_ICSE_2021} tries to capture a finer grained view of expertise by tracking all the APIs that developers have modified. To handle a large dimensional vector of APIs (e.g., 100M dimensional vector), the authors leverage Doc2Vec. However, the approach has two practical drawbacks: (1) including all APIs make it challenging to pinpoint the skills that form the expertise of developers, and (2) building neural network models with vector representation of Doc2Vec for software engineering tasks (e.g., reviewer recommendation \citep{kovalenko2018does}) may result in models that are difficult to deploy and interpret. In summary, \textbf{there exists the need for an approach that is more practical and effective in representing developers' expertise}.

\smallskip \noindent \textbf{Representation of developers' expertise.} Our paper is related to the work of \citet{expert_represent_ICSE_2021}. The authors propose and construct Skill Space, a methodology to represent the expertise of developers by comparing vector representation of four important entities (e.g., developers, projects, programming languages, and APIs). The authors operationalize  Skill Space using the World of Code (WoC) data that contains APIs extracted from source code changes of 17 programming languages. To represent Skill Space, the authors employ Doc2Vec text embedding where each document represents the APIs encountered in a single modified file, and the document tags represent the language, the project, and the developer. The evaluation strategy involves fitting the Doc2Vec model on past data and measuring the alignment (e.g., cosine similarity) between vectors representing any pair of developers, projects, and APIs. The results show that Skill Space representations can help identify important aspects between developers, projects, and APIs. For instance, the authors observe that developers are closely aligned to themselves in the Skill Space when developers use new APIs. The results also show that Skill Space affects the probability of a developer's pull requests being accepted.

Our work is different from the above work. Our approach for constructing developers' expertise profile with KUs is considerably easy to deploy (see Figure~\ref{fig:ku_representation_process} of the Section~\ref{sec:Data_Collection}) whereas deep learning models (e.g., Doc2Vec) are often difficult to deploy and interpret in practice. In addition, we build reviewer recommender with KUs (e.g., KUREC), compare the performance with baselines and demonstrate the usefulness of our KU-based expertise representation in recommending code reviewers for PRs.

%In this work, we define the notion of \textit{KUs} of a programming language and introduce a new perspective for representing developer expertise. We conduct an exploratory study on representing developer expertise with these KUs. Then, we study how knowledge concentration changes with the evolution of software systems. Finally, we study how such representation of developer expertise with KUs can be effective for recommending code reviewers. \goliva{This comparison is not very good. What you say is true, but it is already quite clear that the two approaches vary considerably in their nature. It is probably best to focus on other aspects, such as ease of use/deployment. DL models are often difficult to deploy and interpret in practice. in turn, one can easily derive expertise profiles using KU (refer to the preliminary study and that figure with the 6 profiles/clusters)}.

More generally, several studies have attempted to define metrics for measuring expertise of developers using different sources of information, such as source code changes (e.g., commits and pull requests)~\citep{eval_exp_recom_GROUP_2001,exp_browser_ICSE_2002,dev_drive_soft_evol_IWPSE_2005, assessing_expertise_EASE_2021}, API usage~\citep{usage_expertise_ICSME_2009, agent_to_assist_CHI_2000, Liang22}, source code defects~\citep{expertise_from_bug_report_MSR_2007}, user interaction data~\citep{knowledge_of_code_activity_FSE_2007,degree_of_knowledge_code_familarity_ICSE_2010,degree_of_prog_language_knowledge_TOSEM_2014}, commit messages~\citep{greene2016cvexplorer} and, \texttt{README} files~\citep{wan2018scsminer,hauff2015matching}. We briefly highlight a few of them in the following paragraphs.

\citet{assessing_expertise_EASE_2021} use the changes in syntax patterns of source code as a proxy for measuring programming expertise. The authors propose a model that relies on the Zipf distribution of syntactic patterns in artifacts that are produced by a developer to evaluate developer’s mastery in programming syntax patterns. To evaluate their model's performance, the authors compared the performance of the model with two groups of Clustering and Classifier algorithms. The results show that the proposed approach outperforms the state-of-the-art approaches for classifying novice and expert programmers.

%The authors show that the distribution of syntax patterns follows Zipf's low, and developers' syntactic expertise is reflected in the parameters of the distribution. \goliva{Expand on this one}

\citet{Liang22} present a novel method to detect OSS skills and prototype, namely DISKO (\textbf{D}etect\textbf{I}ng \textbf{SK}ills in \textbf{O}SS). The approach identifies a set of relevant signals, which are measurable activities or cues associated with four skills: (1) teaches others to be involved in the OSS project, (2) shows commitment towards the OSS project, (3) has knowledge in programming languages and (4) has familiarity with OSS practices. The authors evaluate the approach by surveying 455 OSS contributors. The authors observe that their approach can detect the presence of the skills with precision scores between 77\% to 97\%.

Our work is different from the above works because we define KUs that we operationalize using certification exams. These KUs act as proxies for knowledge or skill categories of programming languages. In this paper, we conduct an exploratory study of the expertise representation with KUs. We also demonstrate that such expertise representation with KUs can be effective for recommending code reviewers in collaborative platforms (e.g., GitHub).

%\ahsan{did you forget to remove this paragraph because the detailed of these works are removed? or, we are keeping this paragraph.}%\goliva{keep it on purpose. just to say that there are more things out there.}

%\goliva{Add this paragraph somewhere} Interestingly, similar imbalance (RQ1) has been observed in several other software engineering studies that aimed to classify developers into meaningful groups. Examples include studies in the context of developer activity levels in open source development~\citep{goeminne2011evidence} and in the classification of developers into core and peripheral~\citep{joblin2017classifying}. 

\smallskip \noindent \textbf{Knowledge Units (KUs).} There have been several works on representing knowledge using KUs~\citep{patel2018data,petak2020modelling,li2021towards,fuzzy_ku,cogo2022assessing}. The importance of the hierarchical organization of ontology for an intelligent web is presented by~\citet{patel2018data}. The authors use KUs to represent the semantic web's ontological concepts. They show that KUs can help achieve context-sensitive inference power and acquire new knowledge easily.

\citet{petak2020modelling} construct a fuzzy expert system based on fuzzy KUs which are defined as specific and core production rules for the system. These KUs are used to help decide whether to upgrade an enterprise resource planning system. The results demonstrate that KUs can be useful for these expert systems' learning process. In another study, \citet{fuzzy_ku} demonstrate how complicated issues can be broken down into their most basic forms using a network of fuzzy KUs. 

\citet{von2018identifying} represent the important knowledge for studying system engineers’ cybersecurity models into different KUs (e.g., Software, Component and Connection, System). To ascertain which topics are deemed to be crucial for inclusion in the module, the authors conduct interviews with engineering professionals from academia and industry. According to the interview results, the three most important areas for cybersecurity are software security, systems security, and organizational security.

\citet{cogo2022assessing} encode the Rust domain knowledge into 47 different knowledge units (KUs) to study the topical alignment of the Rust documentation. The authors extract these KUs from both the  Question \& Answer (Q\&A) websites and the official Rust documentation. The authors demonstrate the usefulness of KUs to study the difference between the concrete developers’ information needs and the current state of the documentation.

In this paper, we use KUs as a means to create ``thematic groupings that encompass multiple, related topics \citep{bishop2017cybersecurity}'' of the Java programming language. More specifically, we define a programming language KU as a cohesive set of key capabilities that are offered by one or more building blocks of a given programming language (see Section~\ref{sec:Knowledge_Unit}). We identify Java programming language KUs from the topics and subtopics of the Java certification exams (e.g., the Oracle Java SE and EE certification exams).
    \section{Threats to Validity}
\label{sec:Limitations_And_Threats}

% \goliva{We need to say that we include chRev in lieu of other modern recommenders because the original authors show that it outperforms other models. However, in our results, we observe that both KU and RF outperform chRev. Hence, we suggest that future work should replicate our study and (i) include additional recommender such as REVFINDER or TIE? and (ii) additional subject systems.} \ahsan{added. See the external validity section.}

\smallskip \noindent\textbf{Construct validity.} One of the threats to construct validity is that we did not extract KUs from the third-party libraries that are used by the studied projects. Our objective is to understand the effectiveness of KUs in representing developer expertise. Therefore, we only focus on those KUs that are implemented by the developers of the studied projects. To ensure that we are less likely to miss the KUs that are implemented within the studied projects, we carefully implemented type binding resolutions using the Java JDT library \citep{eclipse_jdt}. 

In this study, we use the certification exams for the Java programming language as guidelines to identify KUs. Our rationale is that certification exams cover important topics which are used to measure the expertise of developers in using the programming language. The set of KUs that we conceive in this paper are based on the core functionalities of the programming language that are documented in the topics of these exams. Therefore, our identified KUs could be limited and might not cover all the knowledge surrounded with the constructs and APIs of the Java programming language. In addition, certification exams of different programming language versions (e.g., Java SE 8 vs Java SE 11) might cover different topics. Our paper should be seen as a first attempt to introduce the notion of KUs and design how to operationalize these KUs. Future research should focus on refining our conceptualization of KUs and operationalizing them in different ways.

We applied PCA to reduce the dimension of our KU expertise representation data before running the k-means clustering algorithm. To reduce the dimensionality of the data, one needs to select a few PCA components. To select these PCA components, we decided to use a threshold-based approach on the explained variance of the PCA components. We selected the first few PCA components that aggregate up to the 95\% of the data which is a reasonable threshold for the explained variance of the data \citep{xibilia2020soft, ghotra_pca_MSR_2017, kondo2019impact}.

\smallskip \noindent\textbf{Internal validity.} There could be a threat in identifying the optimal number of clusters using KUs. Identifying an optimum number of clusters is a non-trivial task. However, we follow the silhouette method that has been widely used in prior works \citep{nidheesh2020hierarchical} for determining the optimal number of clusters. In addition, we carefully calculate the silhouette value using the \code{sklearn.metric} package of Python \citep{silo_sklearn}.

\smallskip \noindent \textbf{External validity.} In this study, we only focus on Java projects of GitHub. We were aware of the fact that GitHub contains projects that are not real-world projects (e.g., academic projects and toy projects). Therefore, we carefully filtered our studied projects using the list of real-world software projects curated by~\citet{engineered_project_github_EMSE_2017}. We encourage future studies to broaden the scope of our study and investigate how our findings apply to software projects written in other popular programming languages (e.g., Python) that have certification exams in different development areas (e.g., web development). For example, popular programming languages for web development (e.g., Ruby on Rails, JavaScript, Python, ASP.Net, Node-js, and Objective-C) have their own certification exams \citep{web_certification_exam}.

To compare the performance of our KUREC to that of the state-of-the-art recommender, we choose the CHREV. We include CHREV in lieu of other modern recommenders because its authors show that CHREV outperforms other recommenders~\citep{thongtanunam2015should,kagdi2008}. However, in our results, we observe that both KUREC and RF outperform CHREV. Hence, we suggest that future work should replicate our study to include (i) additional recommenders such as REVFINDER~\citep{thongtanunam2015should}, TIE~\citep{xia2015should}, and Sofia~\cite{mirsaeedi2020mitigating} and (ii) more subject systems from GitHub.

The results discussed in this paper do not necessarily generalize. We thus encourage future research to (i) use KUs and combine them with other metrics and (ii) evaluate whether our findings generalize to other subject systems that differ substantially from the ones we chose.
    \section{Conclusion}
\label{sec:Conclusion}

% \goliva{TODO: say the idea works and the study should be further expanded to consider not only the progrmaming language but also libraries/APIs}

In this paper, we introduce \textit{Knowledge Units} of programming languages to present developers' expertise and present how KU-based expertise can be useful for recommending code reviewers in PRs. We define a KU as a cohesive set of key capabilities that are offered by one or more building blocks of a given programming language. We operationalize our KUs via certification exams for the Java programming language. As a first paper on the topic, our main objective is to study if KU-based expertise representation (i.e., KUREC recommender) can help recommend code reviewers.

Our results of the study demonstrate that KU-based expertise profiles are useful for recommending code reviewers (RQ1). Combining KUREC with other recommenders resulted in even better recommenders that outperform the baselines (RQ2). And when KUREC deviates from the ground truth, we observe that the recommendations are still frequently reasonable (RQ3). We thus conclude that KUREC and AD\_FREQ are overall superior to the baseline recommenders that we studied. These encouraging results can be helpful to researchers who wish to further study developers' skills or build reviewer recommendation models. We thus invite researchers to (i) explore richer conceptualizations and operationalizations of KUs (e.g., supporting the detection of different design patterns), (ii) evaluate whether our findings generalize to other projects that differ from the ones we chose, (iii) investigate the effectiveness of KUs in solving other practical software engineering tasks that depend on the effective identification and mapping of developers' expertise, such as the development of strategies for increasing the truck factor of a project~\citep{avelino2016novel,ferreira2019algorithms}.
	
	\section*{Data Availability Statement (DAS)}
\label{sec:Data_Availability_Statement}

A supplementary material package is provided online in the following link:
\url{http://www.bit.ly/3bhSFux}. The contents will be made available on a public GitHub
repository once the paper is accepted.
	\section*{Conflict of Interest (COI)}
\label{sec:Conflict_Of_Interest}
The authors declared that they have no conflict of interest.
	
	\begin{footnotesize}
		\bibliographystyle{spbasic}      % basic style, author-year citations
		\bibliography{bib/references.bib}   % name your BibTeX data base
	\end{footnotesize}	

	\clearpage
	
	% Appendix
	\appendix

	\begin{Large}
		\noindent \textbf{Appendix}
	\end{Large}
	
	\normalsize
	\vspace{-1ex}
	\section{Java Certification Exams and Knowledge Units}
\label{appendix:cert-exams}

\subsection{Java SE 8 Programmer I Exam}

Table~\ref{tab:java-se-8-exam-progI} lists the topics and subtopics covered in the \textit{Java SE 8 Programmer I Exam} as per Oracle's official webpage.

\begin{table*}[!htbp]
    \centering
    \caption{Topics and subtopics from the Java SE 8 Programmer I Exam.}
    \label{tab:java-se-8-exam-progI}
    \resizebox{\textwidth}{!}{
    \begin{tabular}{p{3.0cm}p{14cm}}
    \toprule
    \multicolumn{1}{C{3.0cm}}{\textbf{Exam topic}} & \multicolumn{1}{C{11cm}}{\textbf{Exam subtopics}} \\ \midrule
    
    \textbf{[T1]} Java Basics & 
    \textbf{[S1]} Define the scope of variables. \newline
    \textbf{[S2]} Define the structure of a Java class. \newline
    \textbf{[S3]} Create executable Java applications with a main method; run a Java program from the command line; produce console output. \newline
    \textbf{[S4]} Import other Java packages to make them accessible in your code. \newline
    \textbf{[S5]} Compare and contrast the features and components of Java such as: platform independence, object orientation, encapsulation, etc.
    \\ \midrule

    \textbf{[T2]} Working with Data Types & 
    \textbf{[S1]} Declare and initialize variableS(including casting of primitive data types). \newline
    \textbf{[S2]} Differentiate between object reference variables and primitive variables. \newline
    \textbf{[S3]} Know how to read or write to object fields. \newline
    \textbf{[S4]} Explain an object's lifecycle (creation, ``dereference by reassignment'' and garbage collection). 
    \\ \midrule
    
    \textbf{[T3]} Using Operators and Decision Constructs&
    \textbf{[S1]} Use Java operators; use parentheses to override operator precedence. \newline
    \textbf{[S2]} Test equality between strings and other objects using \code{==} and \code{equals()} \newline
    \textbf{[S3]} Create and use \code{if}, \code{if-else}, and ternary constructs \newline
    \textbf{[S4]} Use a \code{switch} statement
    \\ \midrule
    
    \textbf{[T4]} Creating and Using Arrays&
    \textbf{[S1]} Declare, instantiate, initialize and use a one-dimensional array \newline
    \textbf{[S2]} Declare, instantiate, initialize and use a multi-dimensional array
    \\ \midrule
    
    \textbf{[T5]} Using Loop Constructs&
    \textbf{[S1]} Create and use \code{while} loops \newline
    \textbf{[S2]} Create and use \code{for} loops including the enhanced \code{for} loop \newline
    \textbf{[S3]} Create and use \code{do-while} loops \newline
    \textbf{[S4]} Use \code{break} statement \newline
    \textbf{[S5]} Use \code{continue} statement 
    \\ \midrule
    
    \textbf{[T6]} Working Methods and Encapsulation&
    \textbf{[S1]} Create methods with arguments and return values, including overloaded methods. \newline
    \textbf{[S2]} Apply the \code{static} keyword to methods and fields. \newline
    \textbf{[S3]} Create an overloaded method; differentiate between default and user defined constructors. \newline
    \textbf{[S4]} Apply access modifiers. \newline
    \textbf{[S5]} Apply encapsulation principles to a class. \newline
    \textbf{[S6]} Determine the effect upon object references and primitive values when they are passed into methods that change the values.
    \\ \midrule

    \textbf{[T7]} Working with Inheritance &
    \textbf{[S1]} Describe inheritance and its benefits. \newline
    \textbf{[S2]} Develop code that makes use of polymorphism; develop code that overrides methods; differentiate between the type of a reference and the type of an object. \newline
    \textbf{[S3]} Determine when casting is necessary. \newline
    \textbf{[S4]} Use \code{super} and \code{this} to access objects and constructors. \newline
    \textbf{[S5]} Use abstract classes and interfaces..
    \\ \midrule
    
    \textbf{[T8]} Handling Exceptions &
    \textbf{[S1]} Differentiate among checked exceptions, \code{RuntimeException}, and \code{Error}. \newline
    \textbf{[S2]} Create a \code{try-catch} block and determine how exceptions alter normal program flow. \newline
    \textbf{[S3]} Describe the advantages of exception handling. \newline
    \textbf{[S4]} Create and invoke a method that throws an exception. \newline
    \textbf{[S5]} Recognize common exception classes and categorieS(such as \code{NullPointerException}, \code{ArithmeticException}, \code{ArrayIndexOutOfBoundsException}, \code{ClassCastException}).
    \\ \midrule
       
    \textbf{[T9]} Working with Selected classes from the Java API&
    \textbf{[S1]} Manipulate data using the \code{StringBuilder} class and its methods \newline
    \textbf{[S2]} Create and manipulate strings. \newline
    \textbf{[S3]} Create and manipulate calendar data using classes from \code{java.time.LocalDateTime}, \code{java.time.LocalDate}, \code{java.time.LocalTime}, \code{java.time.format.DateTimeFormatter}, \code{java.time.Period}. \newline
    \textbf{[S4]} Declare and use an ArrayList of a given type. \newline
    \textbf{[S5]} Write a simple Lambda expression that consumes a Lambda Predicate expression.

    \\ \midrule 
    \end{tabular}
   }
\end{table*}

\subsection{Java SE 8 Programmer II Exam}
Table~\ref{tab:java-se-8-exam-progII} lists the topics and subtopics covered in the \textit{Java SE 8 Programmer II Exam} as per Oracle's official webpage. 

\begin{table*}[!htbp]
    \centering
    \ssmall
    \caption{Topics and subtopics from the Java SE 8 Programmer II Exam.}
    \label{tab:java-se-8-exam-progII}
    \resizebox{\linewidth}{!}{
    \begin{tabular}{p{3.0cm}p{12cm}}
    \toprule
    \multicolumn{1}{C{3.0cm}}{\textbf{Exam topic}} & \multicolumn{1}{C{12cm}}{\textbf{Exam subtopic}} 
    \\ \midrule
    
    \textbf{[T10]} Java Class Design& 
    \textbf{[S1]} Implement encapsulation \newline
    \textbf{[S2]} Implement inheritance including visibility modifiers and composition \newline
    \textbf{[S3]} Implement polymorphism \newline
    \textbf{[S4]} Override \code{hasCode}, \code{equals}, and \code{toString} methods from \code{Object} class \newline
    \textbf{[S5]} Create and use singleton classes and immutable classes. \newline
    \textbf{[S6]} Develop code that uses the \code{static} keyword on initialize blocks, variables, methods, and classes.
    \\ \midrule
    
    \textbf{[T11]} Advanced Class Design Knowledge Unit& 
    \textbf{[S1]} Develop code that uses abstract classes and methods. \newline
    \textbf{[S2]} Develop code that uses the \code{final}. \newline
    \textbf{[S3]} Create inner classes including static inner classes, local classes, nested classes, and anonymous innter classes. \newline
    \textbf{[S4]} Use enumerated types including methods, and constructors in an \code{enum} type \newline
    \textbf{[S5]} Develop code that declares, implements and/or extends interfaces and use the \code{@Override} annotation. \newline
    \textbf{[S6]} Create and use lambda expressions.
    \\ \midrule
    
    \textbf{[T12]} Generics and Collection Knowledge Unit&
    \textbf{[S1]} Create and use a generic class. \newline
    \textbf{[S2]} Create and use \code{ArrayList}, \code{TreeSet}, \code{TreeMap}, and \code{ArrayDeque} objects. \newline
    \textbf{[S3]} Use \code{java.util.Comparator} and \code{java.lang.Comparable} interfaces. \newline
    \textbf{[S4]} Collections, streams, and filters \newline
    \textbf{[S5]} Iterate using \code{forEach} methods of \code{Streams} and \code{List}. \newline
    \textbf{[S6]} Describe the \code{Stream} interface and the \code{Stream} pipeline. \newline
    \textbf{[S7]} Filter a collection by using lambda expressions. \newline
    \textbf{[S8]} Use method references with streams.
    \\ \midrule
    
    \textbf{[T13]} Lambda Built-in Functional Interfaces &
    \textbf{[S1]} Use the built-in interfaces included in the \code{java.util.function} package such as \code{Predicate}, \code{Consumer}, \code{Function}, and \code{Supplier}. \newline
    \textbf{[S2]} Develop code that uses primitive versions of functional interfaces. \newline
    \textbf{[S3]} Develop code that uses binary versions of functional interfaces. \newline
    \textbf{[S4]} Develop code that uses the \code{UnaryOperator} interface.
    \\ \midrule
    
    \textbf{[T14]} Java Stream API &
    \textbf{[S1]} Develop code to extract data from an object using \code{peek()} and \code{map()} methods including primitive versions of the \code{map()} method \newline
    \textbf{[S2]} Search for data by using search methods of the Stream classes including \code{findFirst}, \code{findAny}, \code{anyMatch}, \code{allMatch}, \code{noneMatch} \newline
    \textbf{[S3]} Develop code that uses the Optional class \newline
    \textbf{[S4]} Develop code that uses Stream data methods and calculation methods \newline
    \textbf{[S5]} Sort a collection using Stream API \newline
    \textbf{[S6]} Save results to a collection using the collect method and group/partition data using the Collectors class \newline
    \textbf{[S7]} Use \code{flatMap()} methods in the Stream API.
    \\ \midrule
    
    \textbf{[T15]} Exceptions and Assertions &
    \textbf{[S1]} Use \code{try-catch} and \code{throws} statements \newline
    \textbf{[S2]} Use \code{catch}, \code{multi-catch}, and \code{finally} clauses \newline
    \textbf{[S3]} Use autoclose resources with a \code{try-with-resources} statement \newline
    \textbf{[S4]} Create custom exceptions and autocloseable resources \newline 
    \textbf{[S5]} Test invariants by using assertions
    \\ \midrule
    
    \textbf{[T16]} Use Java SE 8 Data/Time API &
    \textbf{[S1]} Create and manage date-based and time-based events including a combination of date and time into a single object using \code{LocalDate}, \code{LocalTime}, \code{LocalDateTime}, \code{Instant}, \code{Period}, and \code{Duration} \newline
    \textbf{[S2]} Work with dates and times across timezones and manage changes resulting from daylight savings including Format date and times values. \newline 
    \textbf{[S3]} Define and create and manage date-based and time-based events using \code{Instant}, \code{Period}, \code{Duration}, and \code{TemporalUnit} 
    \\ \midrule

    \textbf{[T17]} Java I/O Fundamentals &
    \textbf{[S1]} Read and write data using the console. \newline
    \textbf{[S2]} Use \code{BufferedReader}, \code{BufferedWriter}, \code{File}, \code{FileReader}, \code{FileWriter}, \code{FileInputStream}, \code{FileOutputStream}, \code{ObjectOutputStream}, \code{ObjectInputStream}, and \code{PrintWriter} in the \code{java.io package}
    \\ \midrule
    
    \textbf{[T18]} Java File I/O (NIO.2)&
    \textbf{[S1]} Use the \code{Path} interface to operate on file and directory paths \newline
    \textbf{[S2]} Use the \code{Files} class to check, read, delete, copy, move, and manage metadata a file or directory \newline
    \textbf{[S3]} Use Stream API with NIO.2 
    \\ \midrule
    
    \textbf{[T19]} Object-Oriented Design Principles&
    \textbf{[S1]} Write code that declares, implements and/or extends interfaces. \newline
    \textbf{[S2]} Choose between interface inheritance and class inheritance \newline
    \textbf{[S3]} Develop code that implements ``is-a'' and/or ``has-a'' relationships. \newline
    \textbf{[S4]} Apply object composition principles. \newline
    \textbf{[S5]} Design a class using the Singleton design pattern. \newline
    \textbf{[S6]} Write code to implement the DAO pattern. \newline
    \textbf{[S7]} Design and create objects using a factory, and use factories from the API.
    \\ \midrule

    \textbf{[T20]} String Processing&
    \textbf{[S1]} Search, parse and build strings \newline
    \textbf{[S2]} Search, parse, and replace strings by using regular expressions. \newline
    \textbf{[S3]} Use string formatting 
    \\ \midrule
    
    \textbf{[T21]} Concurrency&
    \textbf{[S1]} Create worker threads using \code{Runnable}, \code{Callable} and use an \code{ExecutorService} to concurrently execute tasks \newline
    \textbf{[S2]} Identify potential threading problems among deadlock, starvation, livelock, and race conditions. \newline
    \textbf{[S3]} Use \code{synchronized} keyword and \code{java.util.concurrent.atomic} package to control the order of thread execution. \newline
    \textbf{[S4]} Use \code{java.util.concurrent} collections and classes including \code{CyclicBarrier} and \code{CopyOnWriteArrayList}. \newline
    \textbf{[S5]} Use parallel Fork/Join Framework 
    \\ \midrule
    
    \textbf{[T22]} Building Database Applications with JDBC &
    \textbf{[S1]} Describe the interfaces that make up the core of the JDBC API including the \code{Driver}, \code{Connection}, \code{Statement}, and \code{ResultSet} interfaces and their relationship to provider implementations. \newline
    \textbf{[S2]} Identify the components required to connect to a database using the \code{DriverManager} class including the JDBC URL. \newline
    \textbf{[S3]} Submit queries and read results from the database including creating statements, returning result sets, iterating through the results, and properly closing result sets, statements, and connections.
    \\ \midrule    

    \textbf{[T23]} Localization&
    \textbf{[S1]} Read and set the locale by using the \code{Locale} object. \newline
    \textbf{[S2]} Create and read a \code{Properties} file. \newline
    \textbf{[S3]} Build a resource bundle for each locale and load a resource bundle in an application.
    \\ \midrule    
    \end{tabular}
    }
\end{table*}

\begin{table*}[!htbp]
    \centering
    \ssmall
    \caption{Topics and subtopics from the Java EE Developer Exam.}
    \label{tab:java-ee-7-dev-exam}
    \resizebox{\linewidth}{!}{
    \begin{tabular}{p{3.0cm}p{12cm}}
    \toprule
    \multicolumn{1}{C{3.0cm}}{\textbf{Exam topic}} & \multicolumn{1}{C{12cm}}{\textbf{Exam subtopic}} 
    \\ \midrule

    \textbf{[T24]} Understanding Java EE Architecture& 
    \textbf{[S1]} Describe Java EE 7 standards, containers, APIs, and services \newline
    \textbf{[S2]} Differentiate between application component functionalities \newline
    \textbf{[S3]} Demonstrate understanding of Enterprise JavaBeans and CDI beans, their lifecycle and memory scopes \newline
    \textbf{[S4]} Demonstrate understanding of the relationship between bean components, annotations, injections, and JNDI \newline
    \textbf{[S5]} Create and use singleton classes and immutable classes
    \newline
    \textbf{[S6]} Develop code that uses the static keyword on initialize blocks, variables, and methods \\ \midrule

    \textbf{[T25]} Manage Persistence using JPA Entities and BeanValidation&
    \textbf{[S1]} Create JPA Entity and Object-Relational Mappings (ORM) \newline
    \textbf{[S2]} Use Entity Manager to perform database operations, transactions, and locking with JPA entities \newline 
    \textbf{[S3]} Create and execute JPQL statements \\ \midrule

    \textbf{[T26]} Implement Business Logic by Using EJBs&
    \textbf{[S1]} Create session EJB components containing synchronous and asynchronous business methods, manage the life cycle container callbacks, and use interceptors \newline 
    \textbf{[S2]} Create EJB timers \newline 
    \textbf{[S3]} Demonstrate understanding of how to control EJB transactions, distinguish Container Managed (CMT) and Bean Managed (BMT) transactions \\ \midrule 

    \textbf{[T27]} Use Java Message Service API&
    \textbf{[S1]} Implement Java EE message producers and consumers, including Message-Driven beans \newline 
    \textbf{[S2]} Use transactions with JMS API \\ \midrule 

    \textbf{[T28]} Implement SOAP Services by Using JAX-WS and JAXB API&
    \textbf{[S1]} Create SOAP Web Services and Clients using JAX-WS API \newline 
    \textbf{[S2]} Create marshall and unmarshall Java Objects by using JAXB API \\ \midrule 

    \textbf{[T29]} Create Java Web Applications using Servlets&
    \textbf{[S1]} Create Java Servlet and use HTTP methods \newline 
    \textbf{[S2]} Handle HTTP headers, parameters, cookies \newline 
    \textbf{[S3]} Manage servlet life cycle with container callback methods and WebFilters \\ \midrule 

    \textbf{[T30]} Create Java Web Applications using JSPs&
    \textbf{[S1]} Describe JSP life cycle  \newline 
    \textbf{[S2]} Use JSP syntax, use tag libraries and  Expression Language (EL) \newline 
    \textbf{[S3]} Handle errors using Java Server Pages \\ \midrule 
    
    \textbf{[T31]} Implement REST Services using JAX-RS API&
    \textbf{[S1]} Apply REST service conventions \newline 
    \textbf{[S2]} Create REST Services and clients using JAX-RS API \\ \midrule 

    \textbf{[T32]} Create Java Applications using WebSockets&
    \textbf{[S1]} Understand and utilise WebSockets communication style and lifecycle \newline 
    \textbf{[S2]} Create WebSocket Server and Client Endpoint Handlers \newline
    \textbf{[S3]} Produce and consume, encode and decode WebSocket messages \\ \midrule 

    \textbf{[T33]} Develop Web Applications using JSFs&
    \textbf{[S1]} Describe JSF architecture, lifecycle and  navigation \newline 
    \textbf{[S2]} Use JSF syntax and use JSF Tag Libraries \newline
    \textbf{[S3]} Handle localisation and produce messages \newline 
    \textbf{[S4]} Use Expression Language (EL) and interact with CDI beans \\ \midrule 

    \textbf{[T34]} Secure Java EE 7 Applications&
    \textbf{[S1]} Describe Java EE declarative and programmatic security and configure authentication using application roles and security constraints and Login Modules  \newline 
    \textbf{[S2]} Describe WebServices security standards \\ \midrule 

    \textbf{[T35]} Use CDI Beans&
    \textbf{[S1]} Create CDI Bean Qualifiers, Producers, Disposers, Interceptors, Events, and Stereotypes \\ \midrule 

    \textbf{[T36]} Use Concurrency API in Java EE 7 Applications&
    \textbf{[S1]} Demonstrate understanding of Java Concurrency Utilities and use Managed Executors \\ \midrule 

    \textbf{[T37]} Use Batch API in Java EE 7 Applications&
    \textbf{[S1]} Implement batch jobs using JSR 352 API \\ \bottomrule

\end{tabular}
}
\end{table*}

\subsection{Mapping Process From Certification Exams to Knowledge Units (KUs)}
We detail our process for mapping from certification exams to KUs. To map from certification exams to KUs, we follow three steps. First, we exclude subtopics (from either exam) that are unsuitable for our study. Next, we rearrange subtopics (within exams and across exams) such that subtopics lie in the most specific topic possible. Finally, we convert each topic into a KU (one-to-one mapping) and interpret its subtopics as the key capabilities of the KU. Table \ref{tab:java-se-8-exam-progI} and Table \ref{tab:java-se-8-exam-progII} show the topics and subtopics of the Java SE 8 certification exam I (E1) and the Java SE 8 certification exam II (E2), respectively. In the following, we explain the mapping process in detail (we use the notation [Ei, Tj, Sk] to refer to a subtopic Sk from topic Tj of exam Ei).

\smallskip \noindent \textbf{Exclusion of subtopics.} We only considered the subtopics that can be automatically detected from the source code using static analysis. For instance, the subtopic ``Use autoclose resources with a try-with-resources statement'' can be detected by searching for ``try-with-resources'' code blocks using an appropriate Java parser. Our rationale is to ensure that KUs are computable for any Java system, irrespective of how those are built or used by end-users. In this vein, subtopics that are inherently conceptual or that pose great challenges to be detected in practice (e.g., involve problems that are still under investigation by the SE research community) are excluded from our operationalization. We discuss all individual cases below.

The subtopic [E1, T1, S5] under ``[E1, T1] Java Basics'' is about comparing and contrasting the features and components of Java, such as: platform independence, object orientation, encapsulation, etc. This topic is thus more of a concept, which can not be operationalized and extracted automatically. After discussion, the first two authors concluded that none of the subtopics of ``[E1, T1] Java Basics'' can be extracted automatically (e.g., [E1, T1, S5] comparing and contrasting the features and components of Java such as: platform independence, object orientation, encapsulation, etc). As a consequence, we did not include ``[E1, T1] Java Basics'' in our study. 

Some of the subtopics of topics are about the understanding of a concept which can not be operationalized and extracted automatically. For example, the subtopic ``[E1, T7, S1] Describe inheritance and its benefits'' is about understanding the concept of inheritance. The subtopic ``[E1, T8, S3] Describe the advantages of exception handling'' is about understanding the concept of exception handling. The authors did not include such cases of subtopics as capabilities of KUs. We highlight all these cases in Table \ref{tab:map-java-se-8-exam-kus}.

The authors decided not to include the topic ``[E2, T19] Object-Oriented Design Principles'' in our study. The topic ``[E2, T19] Object-Oriented Design Principles'' is about the implementation and detection of different design patterns (e.g., singleton design pattern and factory design pattern). Detection of such design patterns is a non-trivial task and is itself an active research area in the software engineering community \cite{tsantalis2006design,xiong2019accurate}. Hence, the authors did not include the topic ``[E2, T19] Object-Oriented Design Principles'' in this study.

The authors did not include the topic ``[E3, T30] Create Java Web Application using JSPs''. To create Java web application using JSPs, developers need to create a JSP server pages (e.g., files with ``.jsp'' extension) containing the JSP tags which are different from Java files. In this study, we only focus on studying the Java files (e.g., files with ``.java'' extension). Therefore, the authors decided not to include the topic ``[E3, T30]'' in this study.

\noindent \textbf{Rearrangement of subtopics.} Certain subtopics from exam E1 are conceptually related to topics of exam E2 and vice-versa. In those cases, we prioritize the topic that is more specific and move the subtopic in question to that topic. For example, “[E2, T10] Java Class Design” contains the subtopic “[E2, T10, S2] Implement inheritance including visibility modifiers and composition”. However, such a subtopic would also fit under “[E1, T7] Working with Inheritance.” Since the latter topic is more specific than the former, we move [E2, T10, S2] into [E1, T7]. We also observed rare cases in which it made sense to perform subtopic moving within an exam (as opposed to cross exams). In the following, we describe all subtopic rearrangements in detail.

The subtopics of “[E1, T9] Working with selected classes from the Java API” were moved into several topics of E2:
\begin{itemize}
    \item[(1)] The subtopics ``[E1, T9, S1] Manipulate data using the StringBuilder class and its methods'' and ``[E1, T9, S2] Create and manipulate strings'' were moved into ``[E2, T10] String Processing''.
    \item[(2)] The subtopic ``[E1, T9, S3] Create and manipulate calendar data using classes from java.time.LocalDateTime, java.time.LocalDate,java.time.LocalTime, java.time.format.DateTimeFormatter, java.time.Period.'' was moved into ``[E2, T16] Use Java SE 8 Date/Time API''. 
    \item[(3)] The subtopic ``[E1, T9, S4] Declare and use an ArrayList of a given type'' was moved into ``[E2, T12] Generics and Collection''.
    \item[(4)] The subtopic ``[E1, T9, S5] Write a simple Lambda expression that consumes a Lambda Predicate expression'' was moved into ``[E2, T13] Lambda Built-in Functional Interfaces''.
\end{itemize}    

The subtopics ``[E1, T8, S2] Create a try-catch block'' and ``[E1, T8, S4] Create and invoke a method that throws an exception'' were moved from ``[E1.T8] Handling Exception'' into ``[E2.T15] Exceptions and Assertions''.

All subtopics from the topic ``[E2, T10] Java Class Design'' were moved into other topics of E1 and E2 as follows:
\begin{itemize}
    \item[(1)] The subtopics ``[E2, T10, S1] Implement encapsulation'', ``[E2, T10, S4] Override hasCode, equals, and toString methods from Object class'', and ``[E2, T10, S6] Develop code that uses the static keyword on initialize blocks, variables, and methods'' were moved into ``[E1, T6] Working with Methods and Encapsulation''.
    \item[(2)] The subtopics ``[E2, T10, S2] Implement inheritance including visibility modifiers and composition'' and ``[E2, T10, S3] Implement polymorphism'' were moved into ``[E1, T7] Working with Inheritance''.
    \item[(3)] The subtopic ``[E2, T10, S5] Create and use singleton classes and immutable classes'' was moved into ``[E2, T11] Advanced Class Design''.

Three subtopics of the topic ``[E2, T12] Generics and Collection KU'' were moved into ``[E2, T14] Java Stream API''. The subtopics (1) ``[E2, T12, S5] Iterate Collection using forEach methods of Streams'', (2) ``[E2, T12, S7] Filter a collection by using Stream filter API with lambda expressions'', and (3) ``[E2, T12, S8] Use method references with streams'' were moved into ``[E2, T14] Java Stream API''.
\end{itemize}

\noindent \textbf{One-to-one mappings.} After performing the exclusions and subtopic rearrangements, we created a one-to-one mapping between topics and KUs. For example, the topic ``[E1, T5] Using Loop Constructs'' is mapped to the ``[K4] Loop KU''. All of the subtopics in [E1, T5] are interpreted as key capabilities associated with [K4]. For instance, the subtopic ``[E1, T5, S1] Create and use while loops'' is considered a key capability associated with the Loop KU. As a result of our mapping process, we identified 28 KUs, which are shown in Table \ref{tab:map-java-se-8-exam-kus}.

\begin{table*}[!htbp]
    \centering
    \caption{The mapping from certification exams to Knowledge Unit (KUs)}
    \label{tab:map-java-se-8-exam-kus}
    \resizebox{\textwidth}{!}{
    \begin{tabular}
    {   >{\raggedright\arraybackslash}p{1.5cm}%
        >{\raggedright\arraybackslash}p{2.5cm}%
        >{\raggedright\arraybackslash}p{6.5cm}%
        >{\centering\arraybackslash}p{3.0cm}%
        >{\centering\arraybackslash}p{2.5cm}%
        >{\raggedright\arraybackslash}p{4.0cm}%
  }
  \toprule
    \multicolumn{1}{C{1.5cm}}{\textbf{Exam}} & \multicolumn{1}{C{2.5cm}}{\textbf{Topic}} &
    \multicolumn{1}{C{6.5cm}}{\textbf{Subtopic}}&
    \multicolumn{1}{C{3.0cm}}{\textbf{Mapping Action}}&
    \multicolumn{1}{C{2.5cm}}{\textbf{Knowledge Unit (KU)}} &
    \multicolumn{1}{C{4.0cm}}{\textbf{Reason}}\\
    \midrule
    \multirow{5}{1.5cm}{{[E1]} Java SE 8 \\ Programmer \\ I \\ certification \\ exam}  & {[T1]} Java Basics & {[S1]} Define the scope of variables & Excluded & -- & \multirow{2}{4cm}{{[E1, T1]} is basically about the understanding of the concept of the Java programming language and knowledge about how to run a Java program. None of the subtopics of {[E1, T1]} can be operationalized and extracted automatically. So, {[E1, T1]} is excluded from this study.} \\
    & & {[S2]} Define the structure of a Java class & & &  \\
    & & {[S3]} Understand executable Java applications with the main method & & &\\
    & & {[S4]} Understand the process of importing packages to make them accessible in your code & & & \\
    & & {[S5]} Compare and contrast the features and components of Java such as platform independence, object orientation, encapsulation, etc. & & &\\ \cmidrule{2-6}
     & {[T2]} Working with Data Types & {[S1]} Declare and initialize variableS(including casting of primitive data types) & Mapped to KU & {[K1]} Data Type KU  & One to one mapping.  \\ \cmidrule{3-6}
     & & {[S2]} Differentiate between object reference variables and primitive variables & Excluded & -- & \multirow{2}{4cm}{These subtopics of topics are about the understanding of a concept which can not be operationalized and extracted automatically.} \\
    & & {[S3]} Know how to read or write to object fields & & &  \\
    & & {[S4]} Explain an object’s lifecycle (creation, “dereference by reassignment” and garbage collection) & & & \\ \cmidrule{2-6}

    & \multirow{3}{2.5cm}{[T3] Using Operators and Decision Constructs}  & [S1] Use Java operators; use parentheses to override operator precedence & Mapped to KU & \multirow{3}{2cm}{[K2] Operator and Decision KU} & One to one mapping.  \\
    & & [S2] Test equality between strings and other objects using ``='' and ``equals( )''. & & & \\
    & & [S3] Create and use if, if-else, and ternary constructs & & & \\
    & & [S4] Use a switch statement  & & & \\ \cmidrule{2-6}
    
    & [T4] Creating and Using Arrays & [S1] Declare, instantiate, initialize and use a one-dimensional array & Mapped to KU & [K3] Array KU & One to one mapping. \\
    & & [S2] Declare, instantiate, initialize and use a multi-dimensional array & & & \\ \cmidrule{2-6}
    
    & \multirow{2}{2.5cm}{[T5] Using Loop Constructs} & [S1] Create and use "while" loops.  & Mapped to KU & [K4] Loop KU & One to one mapping. \\
    & & [S2] Create and use for loops, including the "enhanced for" loop & & & \\
    & & [S3] Create and use do-while loops & & & \\
    & & [S4] Use break statement & & & \\
    & & [S5] Use continue statement & & & \\ \cmidrule{2-6}
    
    & \multirow{4}{2.5cm}{[T6] Working with Methods and Encapsulation} & [S1] Create methods with arguments and return values, including overloaded methods & Mapped to KU & \multirow{4}{2cm}{[K5] Method and Encapsulation KU} & One to one mapping. \\
    & & [S2] Apply the static keyword to methods and fields & & & \\
    & & [S3] Create an overloaded method; create default and user-defined constructors & & & \\
    & & [S3] Create an overloaded method; create default and user-defined constructors & & & \\
    & & [S5] Apply encapsulation principles to a class & & & \\ \cmidrule{3-6}
    & & [S6] Determine the effect upon object references and primitive values when they are passed into methods that change the values & Excluded & -- & This subtopic [E1, T6, S6] is about the understanding of a concept which can not be operationalized and extracted automatically. \\ \cmidrule{3-6}
    & & [S7] Create a constructor chaining (use“this()” method to call one constructor from another constructor & Mapped to KU & \multirow{4}{2cm}{[K5] Method and Encapsulation KU} & One to one mapping. \\
    & & [S8] Use variable length arguments in the methods & & & \\ \cmidrule{2-6}
    
    & [T7] Working with Inheritance & [S1] Describe inheritance and its benefits & Excluded & -- & This subtopic [E1, T7, S1] is about the understanding of a concept which can not be operationalized and extracted automatically. \\ \cmidrule{3-6}
    & & [S2] Develop code that makes use of polymorphism; develop code that overrides methods; differentiate between the type of a reference and the type of an object & Mapped to KU & [K6] Inheritance KU & One to one mapping. \\
    & & [S3] Use casting to subclass  & & & \\
    & & [S4] Use super and this to access objects and constructors & & & \\
    & & [S5] Use abstract classes and interfaces & & & \\ \cmidrule{2-6}
    
    & [T8] Handling Exceptions & [S1] Differentiate among checked exceptions, RuntimeException, and Error & Excluded & -- & The subtopic [E1, T8, S1] is about the understanding of a concept which can not be operationalized and extracted automatically.\\ \cmidrule{3-6}
    & & [S2] Create a try-catch block and determine how exceptions alter normal program flow & Moved into [E2, T15] Exceptions and Assertions & [K11] Exception and Assertion KU & Avoid conceptual overlap between the two certification exams.\\ \cmidrule{3-6}
    & & [S3] Describe the advantages of exception handling & Excluded & -- & The subtopic [E1, T8, S3] is about the understanding of a concept which can not be operationalized and extracted automatically. \\ \cmidrule{3-6}
    
    \end{tabular}
   }
\end{table*}

\begin{table*}[!htbp]
    \centering
    %\caption{The mapping from certification exams to Knowledge UnitS(KUs)}
    %\label{tab:java-se-8-exam-progI}
    \resizebox{\textwidth}{!}{
    \begin{tabular}
    {   >{\raggedright\arraybackslash}p{1.5cm}%
        >{\raggedright\arraybackslash}p{2.5cm}%
        >{\raggedright\arraybackslash}p{6.5cm}%
        >{\centering\arraybackslash}p{3.0cm}%
        >{\centering\arraybackslash}p{2.5cm}%
        >{\raggedright\arraybackslash}p{4.0cm}%
  }
  \toprule
    \multicolumn{1}{C{1.5cm}}{\textbf{Exam}} & \multicolumn{1}{C{2.5cm}}{\textbf{Topic}} &
    \multicolumn{1}{C{6.5cm}}{\textbf{Subtopic}}&
    \multicolumn{1}{C{3.0cm}}{\textbf{Mapping Action}}&
    \multicolumn{1}{C{2.5cm}}{\textbf{Knowledge Unit (KU)}} &
    \multicolumn{1}{C{4.0cm}}{\textbf{Reason}}\\
    \midrule
    & & [S4] Create and invoke a method that throws an exception & \multirow{3}{3cm}{Moved into [E2, T15] Exceptions and Assertions} & [K11] Exception KU & \multirow{3}{4cm}{Avoid conceptual overlap between the two certification exams.}\\
    & & [S5] Use common exception classes and categories(such as NullPointerException, ArithmeticException, ArrayIndexOutOfBoundsException, ClassCastException) & & &\\ \cmidrule{2-6}

    & [T9] Working with selected classes from the Java API & [S1] Manipulate data using the StringBuilder class and its methods & Moved into [E2, T20] String Processing & [K15] String Processing KU & Avoid conceptual overlap between the two certification exams. \\
    & & [S2] Create and manipulate strings & & &\\ \cmidrule{3-5}
    & & [S3] Create and manipulate calendar data using classes from java.time.LocalDateTime, java.time.LocalDate,java.time.LocalTime, java.time.format.DateTimeFormatter, java.time.Period & Moved into [E2, T16] Use Java SE 8 Date/Time API & [K12] Date Time API KU &\\ \cmidrule{3-5}
    & & [S4] Declare and use an ArrayList of a given type & Moved into [E2, T12] Generics and Collection & [K8] Generics and Collection KU &\\ \cmidrule{3-5}
    & & [S5] Write a simple Lambda expression that consumes a Lambda Predicate expression & Moved into [E2, T13] Lambda Built-in Functional Interfaces & [K9] Functional Interface KU &\\ \midrule

    \multirow{3}{1.5cm}{[E2] Java SE 8 Programmer II certification exam}  & [T10] Java Class Design & [S1] Implement encapsulation & Moved into [E1, T6] Working with Methods and Encapsulation & [K5] Method and Encapsulation KU & Avoid conceptual overlap between the two certification exams. \\ \cmidrule{3-5}
    & & S2] Implement inheritance including visibility modifiers and composition & Moved into [E1, T7] Working with Inheritance & [K6] Inheritance KU & \\
    & & [S3] Implement polymorphism & & & \\ \cmidrule{3-5}
    & & [S4] Override hasCode, equals, and toString methods from Object class & Moved into [E1, T6] Working with Methods and Encapsulation & [K5] Method and Encapsulation KU & \\ \cmidrule{3-6}
    & & [S5] Create and use singleton classes and immutable classes & Moved into [E2, T11] Advanced Class Design & [K7] Advanced Class Design KU & Avoid conceptual overlap between the topics of the certification exams. \\ \cmidrule{3-6}
    & & [S6] Develop code that uses the static keyword on initialize blocks, variables, and methods & Moved into [E1, T6] Working with Methods and Encapsulation & [K5] Method and Encapsulation KU & Avoid conceptual overlap between the two certification exams \\ \cmidrule{2-6}

    & [T11] Advanced Class Design & [S1] Develop code that uses abstract classes and methods & Moved into [E1, T7] Working with Inheritance & [K6] Inheritance KU & Avoid conceptual overlap between the two certification exams. \\ \cmidrule{3-6}
    & & [S2] Develop code that uses the final classes & Mapped to KU & \multirow{3}{2cm}{[K7] Advanced Class Design KU} & One to one mapping. \\
    & & [S3] Create inner classes, including static inner classes, local classes, nested classes, and anonymous inner classes & & & \\
    & & [S4] Use enumerated types including methods and constructors in an enum type & & & \\ 
    & & [S5] Develop code that uses the @Override annotation & & & \\
    & & [S6] Create and use lambda expressions & & & \\ \cmidrule{2-6}
    & \multirow{2}{2.5cm}{[T12] Generics and Collection} & [S1] Create and use a generic class & Mapped to KU & \multirow{2}{2.5cm}{[K8] Generics and Collection KU} & One to one mapping. \\
    & & [S2] Create and useArrayList, TreeSet, TreeMap, and ArrayDequeobjects & & & \\
    & & [S3] Use java.util.Comparator and java.lang.Comparable interfaces & & & \\
    & & [S4] Iterate using forEach methods of List & & & \\ \cmidrule{3-6}
    & & [S5] Describe and use the Stream interface and the Stream pipeline & \multirow{2}{3cm}{Moved into [E2, T14] Java Stream API} & \multirow{2}{2.5cm}{[K10] Stream API KU} & \multirow{3}{4cm}{Avoid conceptual overlap between the topics of the certification exams.} \\
    & & [S6] Filter a collection by using lambda expressions & & & \\
    & & [S7] Use method references with streams & & & \\ \cmidrule{2-6}
    
    & [T13] Built-in Functional Interfaces & [S1] Use the built-in interfaces included in the java.util.function package such as Predicate, Consumer, Function, and Supplier & Mapped to KU & [K9] Functional Interface KU & One to one mapping. \\
    & & [S2] Develop code that uses primitive versions of functional interfaces & & & \\
    & & [S3] Develop code that uses binary versions of functional interfaces & & & \\
    & & [S4] Develop code that uses theUnaryOperatorinterface & & & \\ \cmidrule{2-6}
    
    & [T14] Java Stream API & [S1] Develop code to extract data from an object using peek() and map() methods, including primitive versions of the map() method & Mapped to KU & [K10] Stream API KU & One to one mapping. \\
    & & [S2] Search for data by using search methods of the Stream classes, including findFirst, findAny, anyMatch, allMatch, noneMatch & & & \\
    & & [S3] Develop code that uses the Optional class & & & \\
    & & [S4] Develop code that uses Stream data methods and calculation methods & & & \\
    & & [S5] Sort a collection using Stream API & & & \\
    & & [S6] Save results to a collection using the collect method and group/partition data using the Collectors class & & & \\
    & & [S7] Use flatMap() methods in the Stream API & & & \\
    \midrule
    \end{tabular}
   }
\end{table*}

\begin{table*}[!htbp]
    \centering
    %\label{tab:java-se-8-exam-progI}
    \resizebox{\textwidth}{!}{
    \begin{tabular}
    {   >{\raggedright\arraybackslash}p{1.5cm}%
        >{\raggedright\arraybackslash}p{2.5cm}%
        >{\raggedright\arraybackslash}p{6.5cm}%
        >{\centering\arraybackslash}p{3.0cm}%
        >{\centering\arraybackslash}p{2.5cm}%
        >{\raggedright\arraybackslash}p{4.0cm}%
  }
  \toprule
    \multicolumn{1}{C{1.5cm}}{\textbf{Exam}} & \multicolumn{1}{C{2.5cm}}{\textbf{Topic}} &
    \multicolumn{1}{C{6.5cm}}{\textbf{Subtopic}}&
    \multicolumn{1}{C{3.0cm}}{\textbf{Mapping Action}}&
    \multicolumn{1}{C{2.5cm}}{\textbf{Knowledge Unit (KU)}} &
    \multicolumn{1}{C{4.0cm}}{\textbf{Reason}}\\ \midrule
    &  \multirow{2}{2.5cm}{[T15] Exceptions and Assertions} & [S1] Use try-catch and throws statements & Mapped to KU & \multirow{2}{2.5cm}{[K11] Exception KU} & One to one mapping. \\
    & & [S2] Use catch, multi-catch, and finally clauses & & & \\
    & & [S3] Use autoclose resources with a try-with-resources statement & & & \\
    & & [S4] Create custom exceptions and autocloseable resources & & & \\
    & & [S5]Test invariants by using assertions & & & \\ \cmidrule{2-6}
    
    & [T16] Use Java SE 8 Data/Time API & [S1] Create and manage date-based and time-based events, including a combination of date and time into a single object using LocalDate, LocalTime, LocalDateTime, Instant, Period, and Duration & Mapped to KU & [K12] Date Time API KU & One to one mapping. \\
    & & [S2] Work with dates and times across timezones and manage changes resulting from daylight savings, including Format date and times values & & & \\
    & & [S3] Define and create and manage date-based and time-based events using Instant, Period, Duration, and Temporal Unit & & & \\ \cmidrule{2-6}
    
    & \multirow{2}{1.5cm}{[T17] Java I/O Fundamentals} & [S1] Read and write data using the console & Mapped to KU & [K13] IO KU & One to one mapping. \\
    & & [S2] Use BufferedReader, BufferedWriter, File, FileReader, FileWriter, FileInputStream, FileOutputStream, ObjectOutputStream, ObjectInputStream, and PrintWriter in the java.io package & & & \\ \cmidrule{2-6}
    
    & [T18] Java File I/O (NIO.2) & [S1] Use the Path interface to operate on file and directory paths & Mapped to KU & [K14] NIO KU & One to one mapping. \\
    & & [S2] Use the Files class to check, read, delete, copy, move, and manage metadata of a file or directory & & & \\ \cmidrule{2-6}
    
    & \multirow{3}{1.5cm}{[T19] Object-Oriented Design Principles} & [S1] Develop code that implements ``is-a'' and/or ``has-a'' relationships & Excluded & -- & \multirow{4}{4cm}{This topic is about the implementation and detection of different design patterns. This topic is not included in this study because detection of such design patterns is itself an active research in the software engineering community \cite{tsantalis2006design,xiong2019accurate}.} \\
    & & [S2] Apply object composition principles & & & \\
    & & [S3] Design a class using the Singleton design pattern & & & \\
    & & [S4] Write code to implement the DAO pattern & & & \\
    & & [S5] Design and create objects using a factory, and use factories from the API \newline & & & \\ \cmidrule{2-6}
    
    & \multirow{2}{1.5cm}{[T20] String Processing} & [S1] Search, parse and build strings & Mapped to KU & \multirow{2}{2.5cm}{[K15] String Processing KU} & One to one mapping. \\
    & & [S2] Search, parse, and replace strings by using regular expressions & & & \\
    & & [S3] Use string formatting & & & \\ \cmidrule{2-6}
    
    & [T21] Concurrency & [S1] Create worker threads using Runnable, Callable and use an ExecutorService to concurrently execute tasks & Mapped to KU & [K16] Concurrency KU & One to one mapping. \\
    & & [S2] Identify potential threading problems among deadlock, starvation, livelock, and race conditions & & & \\
    & & [S3] Use synchronized keyword and java.util.concurrent.atomic package to control the order of thread execution & & & \\
    & & [S4] Use java.util.concurrent collections and classes including CyclicBarrierandCopyOnWriteArrayList & & & \\ \cmidrule{2-6}
    
    & \multirow{4}{2.5cm}{[T22] Building Database Applications with JDBC} & [S5] Use parallel Fork/Join Framework & Mapped to KU & [K17] Database KU & One to one mapping. \\ 
    & & [S1] Use the interfaces that make up the core of the JDBC API including the Driver, Connection, Statement, and ResultSet interfaces & & & \\
    & & [S2] Identify the components required to connect to a database using the DriverManagerclass including the JDBC URL &  &  & \\
    & & [S3] Submit queries and read results from the database, including creating statements, returning result sets, iterating through the results, and properly closing result sets, statements, and connections & & & \\ \cmidrule{2-6}
    
    & [T23] Localization & [S1] Read and set the locale by using the Locale object & Mapped to KU & [K18] Localization Knowledge & One to one mapping. \\
    & & [S2] Create and read a Properties file & & & \\
    & & [S3] Build a resource bundle for each locale and load a resource bundle in an application & & & \\ \midrule

    \multirow{5}{1.5cm}{[E3] Java EE certification exam}  & \multirow{4}{2.5cm}{[T24] Understanding Java EE Architecture} & [S1] Describe Java EE 7 standards, containers, APIs, and services & Excluded & -- & \multirow{2}{4cm}{{[E3, T24]} is basically about the understanding of the concept of the Java EE architecture and components. None of the subtopics of {[E3, T24]} can be operationalized and extracted automatically. So, {[E3, T24]} is excluded from this study.} \\ 
    & & [S2] Differentiate between application component functionalities &  &  & \\
    & & [S3] Demonstrate understanding of Enterprise JavaBeans and CDI beans, their lifecycle and memory scopes & & & \\ 
    & & [S4] Demonstrate understanding of the relationship between bean components, annotations, injections, and JNDI &  & \\ 
    & & [S5] Create and use singleton classes and immutable classes &  &  \\ 
    & & [S6] Develop code that uses the static keyword on initialize blocks, variables, and methods & & & \\ \cmidrule{2-6}
    
    & \multirow{4}{2.5cm}{[T25] Manage Persistence using JPA Entities and BeanValidation} & [S1] Create JPA Entity and Object-Relational Mappings (ORM) & Mapped to KU & [K19] Java Persistence KU & One to one mapping. \\

    & & c & & &\\

    & & [S3] Create and execute JPQL statements & & & \\ \cmidrule{2-6}
    
    \end{tabular}
   }
\end{table*}

\begin{table*}[!htbp]
    \centering
    %\label{tab:java-se-8-exam-progI}
    \resizebox{\textwidth}{!}{
    \begin{tabular}
    {   >{\raggedright\arraybackslash}p{1.5cm}%
        >{\raggedright\arraybackslash}p{2.5cm}%
        >{\raggedright\arraybackslash}p{6.5cm}%
        >{\centering\arraybackslash}p{3.0cm}%
        >{\centering\arraybackslash}p{2.5cm}%
        >{\raggedright\arraybackslash}p{4.0cm}%
  }
  \toprule
    \multicolumn{1}{C{1.5cm}}{\textbf{Exam}} & \multicolumn{1}{C{2.5cm}}{\textbf{Topic}} &
    \multicolumn{1}{C{6.5cm}}{\textbf{Subtopic}}&
    \multicolumn{1}{C{3.0cm}}{\textbf{Mapping Action}}&
    \multicolumn{1}{C{2.5cm}}{\textbf{Knowledge Unit (KU)}} &
    \multicolumn{1}{C{4.0cm}}{\textbf{Reason}}\\ \midrule

    & [T26] Implement Business Logic by Using EJBs & [S1] Create session EJB components containing synchronous and asynchronous business methods, manage the life cycle container callbacks, and use interceptors & Mapped to KU & [K20] Enterprise Java Bean KU & One to one mapping. \\

    & & [S2] Create EJB timers & & & \\ \cmidrule{3-6} 

    & & [S3] Demonstrate understanding of how to control EJB transactions, distinguish Container Managed (CMT) and Bean Managed (BMT) transactions & Excluded & -- & This subtopic [E3, T26, S3] is about the understanding of a concept which can not be operationalized and extracted automatically. \\ \cmidrule{2-6} 

    & \multirow{2}{2.5cm}{[T27] Use Java Message Service API} & [S1] Implement Java EE message producers and consumers, including Message-Driven beans & 
    Mapped to KU & [K21] Java Message Service API KU & One to one mapping. \\ 
    & & [S2] Use transactions with JMS API & & & \\ \cmidrule{2-6} 

    & \multirow{4}{2.5cm}{[T28] Implement SOAP Services by Using JAX-WS and JAXB API} & [S1] Create SOAP Web Services and Clients using JAX-WS API & Mapped to KU & [K22] SOAP Web Service KU & One to one mapping. \\ 
    & & [S2] Create marshall and unmarshall Java Objects by using JAXB API & & & \\ \cmidrule{2-6}

    & \multirow{4}{2.5cm}{[T29] Create Java Web Applications using Servlets} & [S1] Create Java Servlet and use HTTP methods & Mapped to KU & [K23] Servlet KU  & One to one mapping. \\ 
    & & [S2] Handle HTTP headers, parameters, cookies & & & \\
    & & [S3] Manage servlet life cycle with container callback methods and WebFilters \\ \cmidrule{2-6}

    & \multirow{4}{2.5cm}{[T30] Create Java Web Applications using JSPs} &  [S1] Describe JSP life cycle & Excluded & --  & \multirow{4}{4cm}{[E3,T30] is excluded as we parse only .java files not .jsp for this study.} \\ 
    & & [S2] Use JSP syntax, use tag libraries and  Expression Language (EL)  & & & \\
    & & [S3] Handle errors using Java Server Pages & & & \\ \cmidrule{2-6}

    & \multirow{3}{2.5cm}{[T31] Implement REST Services using JAX-RS API} & [S1] Apply REST service conventions & Mapped to KU & \multirow{2}{2.5cm}{[K24] Java REST API KU}  & One to one mapping. \\ 
    & & [S2] Create REST Services and clients using JAX-RS API & & & \\ \cmidrule{2-6}

    & \multirow{3}{2.5cm}{[T32] Create Java Applications using WebSockets} & [S1] Understand and utilise WebSockets communication style and lifecycle & Excluded & -- & This subtopic [E3, T32, S1] is about the understanding of a concept which can not be operationalized and extracted automatically. \\ \cmidrule{3-6}

    & & [S2] Create WebSocket Server and Client Endpoint Handlers & Mapped to KU & [K25] Websocket KU  & One to one mapping. \\
    & & [S3] Produce and consume, encode and decode WebSocket messages & & & \\ \cmidrule{2-6}

    &  \multirow{3}{2.5cm}{[T33] Develop Web Applications using JSFs} & [S1] Describe JSF architecture, lifecycle and  navigation & Excluded & -- & This subtopic [E3, T33, S1] is about the understanding of a concept which can not be operationalized and extracted automatically. \\ \cmidrule{3-6}
    & & [S2] Use JSF syntax and use JSF Tag Libraries & Mapped to KU & \multirow{2}{2.5cm}{[K26] Java Server Faces KU} & One to one mapping. \\
    & & [S3] Handle localisation and produce messages & & & \\
    & & [S4] Use Expression Language (EL) and interact with CDI beans & & & \\ \cmidrule{2-6}

    & [T34] Secure Java EE 7 Applications & [S1] Describe Java EE declarative and programmatic security and configure authentication using application roles and security constraints and Login Modules & Excluded & -- & \multirow{5}{4cm}{This topic [E3, T34] is about the understanding of a concept which can not be operationalized and extracted automatically.} \\
    & & [S2] Describe WebServices security standards & & & \\ \cmidrule{2-6}

    & \multirow{3}{2.5cm}{[T35] Use CDI Beans} & [S1] Create CDI Bean Qualifiers, Producers, Disposers, Interceptors, Events, and Stereotypes \newline & Mapped to KU & \multirow{3}{2.5cm}{[K27] Contexts and Dependency Injection (CDI) KU} & One to one mapping. \\ \cmidrule{2-6}
    
    & [T36] Use Concurrency API in Java EE 7 Applications & [S1] Demonstrate understanding of Java Concurrency Utilities and use Managed Executors & Moved into [E2, T21] Concurrency & [K16] Concurrency KU & Avoid conceptual overlap between the two certification exams.\\ \cmidrule{2-6}

    & [T37] Use Batch API in Java EE 7 Applications & [S1] Implement batch jobs using JSR 352 API & Mapped to KU & [K28] Batch Processing KU & One to one mapping. \\

    \bottomrule
    \multirow{1}{20cm}{\pbox{20cm}{* The notation [Ei, Tj, Sk] is employed to refer to a subtopic Sk from topic Tj of exam Ei}} 
\end{tabular}
}
\end{table*}

\begin{table*}[!htbp]
    \centering
    \caption{Knowledge Units derived from the Java SE 8 Programmer I Exam, Java SE 8 Programmer II Exam, and Java EE Developer Exam}
    \label{tab:ku-from-java-exams}
    \resizebox{\textwidth}{!}{
    \begin{tabular}{p{3.0cm}p{16cm}}
    \toprule
    \multicolumn{1}{C{3.0cm}}{\textbf{Knowledge Unit (KU)}} & \multicolumn{1}{C{16cm}}{\textbf{Key Capabilities}} \\ \midrule
    
    \textbf{[K1]} Data Type KU & 
    \textbf{[C1]} Declare and initialize different types of variableS(e.g., primitive type, parameterized type, and array type), including the casting of primitive data types
    \\ \midrule
    \textbf{[K2]} Operator and Decision KU&
    \textbf{[C1]} Use Java operators(e.g., assignment and postfix operators); use parentheses to override operator precedence \newline
    \textbf{[C2]} Test equality between strings and other objects using \texttt{==} and \texttt{equals()} \newline
    \textbf{[C3]} Create and use \texttt{if}, \texttt{if-else}, and ternary constructs \newline
    \textbf{[C4]} Use a \texttt{switch} statement
    \\ \midrule
    
    \textbf{[K3]} Array KU&
    \textbf{[C1]} Declare, instantiate, initialize and use a one-dimensional array \newline
    \textbf{[C2]} Declare, instantiate, initialize and use a multi-dimensional array 
    \\ \midrule
    
    \textbf{[K4]} Loop KU&
    \textbf{[C1]} Create and use \texttt{while} loops \newline
    \textbf{[C2]} Create and use \texttt{for} loops, including the \texttt{enhanced for} loop \newline
    \textbf{[C3]} Create and use \texttt{do-while} loops \newline
    \textbf{[C4]} Use \texttt{break} statement \newline
    \textbf{[C5]} Use \texttt{continue} statement
    \\ \midrule
    
    \textbf{[K5]} Method and Encapsulation KU&
    \textbf{[C1]} Create methods with arguments and return values \newline
    \textbf{[C2]} Apply the ``static'' keyword to methods, fields, and blocks \newline
    \textbf{[C3]} Create an overloaded method and overloaded constructor \newline
    \textbf{[C4]} Create a constructor chaining (use ``this()'' method to call one constructor from another constructor \newline
    \textbf{[C5]} Use variable length arguments in the methods \newline
    \textbf{[C6]} Use different access modifiers (e.g., private and protected) other than ``default''
    \textbf{[C7]} Apply encapsulation: identify set and get method to initialize any private class variables \newline
    \textbf{[C8]} Apply encapsulation: Immutable class generation-final class and initialize private variables through the constructor
    \\ \midrule
    
    \textbf{[K6]} Inheritance KU&
    \textbf{[C1]} Use basic polymorphis (e.g., a superclass refers to a subclass) \newline
    \textbf{[C2] }Use polymorphic parameter (e.g., pass instances of a subclass or interface to a method) \newline
    \textbf{[C3]} Create overridden methods \newline
    \textbf{[C4]} Create ``abstract'' classes and ``abstract'' methods \newline
    \textbf{[C5]} Use ``super()'' and the ``super'' keyword to access the members(e.g., fields and methods) of a parent class \newline
    \textbf{[C6]} Use casting in referring a subclass object to a superclass object
    \\ \midrule
    
    \textbf{[K7]} Advanced Class Design KU&
    \textbf{[C1]} Create inner classes, including static inner classes, local classes, nested classes, and anonymous inner classes \newline
    \textbf{[C2]} Develop code that uses the final \newline
    \textbf{[C3]} Use enumerated types including methods and constructors in an ``enum'' type \newline
    \textbf{[C4]} Create singleton classes and immutable classes
    \\ \midrule 
    
    \textbf{[K8]} Generics and Collection KU&
    \textbf{[C1]} Create and use a generic class \newline
    \textbf{[C2]} Create and use \texttt{ArrayList}, \texttt{TreeSet}, \texttt{TreeMap}, and \texttt{ArrayDeque} \newline
    \textbf{[C3]} Use \texttt{java.util.Comparator} and \texttt{java.lang.Comparable} interfaces \newline
    \textbf{[C4]} Iterate using forEach methods of List
    \\ \midrule

    \textbf{[K9]} Functional Interface KU&
    \textbf{[C1]} Use the built-in interfaces included in the \texttt{java.util.function} packages such as \texttt{Predicate}, \texttt{Consumer}, \texttt{Function}, and \texttt{Supplier} \newline
    \textbf{[C2]} Develop code that uses primitive versions of functional interfaces \newline
    \textbf{[C3]} Develop code that uses binary versions of functional interfaces \newline
    \textbf{[C4]} Develop code that uses the \texttt{UnaryOperator} interface
    \\ \midrule
    
    \textbf{[K10]} Stream API KU&
    \textbf{[C1]} Develop code to extract data from an object using \texttt{peek()} and \texttt{map()} methods, including primitive versions of the \texttt{map()} method \newline
    \textbf{[C2]} Search for data by using search methods of the Stream classes, including \texttt{findFirst}, \texttt{findAny}, \texttt{anyMatch}, \texttt{allMatch}, \texttt{noneMatch} \newline
    \textbf{[C3]} Develop code that uses the Optional class \newline
    \textbf{[C4]} Develop code that uses Stream data methods and calculation methods \newline
    \textbf{[C5]} Sort a collection using Stream API \newline
    \textbf{[C6]} Save results to a collection using the collect method \newline
    \textbf{[C7]} \texttt{UseflatMap()} methods in the Stream API
    \\ \midrule
    
    \textbf{[K11]} Exception KU&
    \textbf{[C1]} Create a try-catch block \newline
    \textbf{[C2]} Use catch, multi-catch, and finally clauses \newline
    \textbf{[C3]} Use autoclose resources with a try-with-resources statement \newline
    \textbf{[C4]} Create custom exceptions and autocloseable resources \newline
    \textbf{[C5]} Create and invoke a method that throws an exception \newline
    \textbf{[C6]} Use common exception classes and categories(such as \texttt{NullPointerException}, \texttt{ArithmeticException}, \texttt{ArrayIndexOutOfBoundsException}, \texttt{ClassCastException}) \newline
    \textbf{[C5]} Use assertions
    \\ \midrule
    
    \textbf{[K12]} Date Time API KU&
    \textbf{[C1]} Create and manage date-based and time-based events including a combination of date and time into a single object using \texttt{LocalDate}, \texttt{LocalTime}, \texttt{LocalDateTime}, \texttt{Instant}, \texttt{Period}, and \texttt{Duration} \newline
    \textbf{[C2]} Formatting date and times values for using different timezones \newline
    \textbf{[C3]} Create and manage date-based and time-based events using Instant, Period, Duration, and Temporal Unit \newline
    \textbf{[C4]} Create and manipulate calendar data using classes from \texttt{java.time.LocalDateTime}, \texttt{java.time.LocalDate}, \texttt{java.time.LocalTime}, \texttt{java.time.format.DateTimeFormatter}, and \texttt{java.time.Period} 
    \\ \midrule
    
    \textbf{[K13]} IO KU&
    \textbf{[C1]} Read and write data using the console \newline
    \textbf{[C2]} Use \texttt{BufferedReader}, \texttt{BufferedWriter}, \texttt{File}, \texttt{FileReader}, \texttt{FileWriter}, \texttt{FileInputStream}, \texttt{FileOutputStream}, \texttt{ObjectOutputStream}, \texttt{ObjectInputStream}, and \texttt{PrintWriter} in the \texttt{java.io} package
    \\ \midrule
    
    \textbf{[K14]} NIO KU&
    \textbf{[C1]} Use the Path interface to operate on file and directory paths \newline
    \textbf{[C2]} Use the Files class to check, read, delete, copy, move, and manage metadata a file or directory
    \\ \midrule
    
    \textbf{[K15]} String Processing KU&
    \textbf{[C1]} Search, parse and build strings \newline
    \textbf{[C2]} Manipulate data using the \texttt{StringBuilder} class and its methods \newline
    \textbf{[C3]} Use regular expression using the \texttt{Pattern} and \texttt{Matcher} class \newline
    \textbf{[C4]} Use string formatting
    \\ \midrule
    \end{tabular}
   }
\end{table*}

\begin{table*}[!htbp]
    \centering
    \resizebox{\textwidth}{!}{
    \begin{tabular}{p{4.0cm}p{15cm}}
    \toprule
    \multicolumn{1}{C{3.0cm}}{\textbf{Knowledge Unit (KU)}} & \multicolumn{1}{C{11cm}}{\textbf{Key Capabilities}} \\ \midrule
    
    \textbf{[K16]} Concurrency KU&
    \textbf{[C1]} Create worker threads using \texttt{Runnable}, \texttt{Callable} and use an \texttt{ExecutorService} to concurrently execute tasks
    \textbf{[C2]} Use \texttt{synchronized} keyword and \texttt{java.util.concurrent.atomic} package to control the order of thread execution \newline
    \textbf{[C3]} Use \texttt{java.util.concurrent}   collections and classes including \texttt{CyclicBarrier} and \texttt{CopyOnWriteArrayList} \newline
    \textbf{[C4]} Use parallel Fork/Join Framework
    \\ \midrule
    
    \textbf{[K17]} Database KU&
    \textbf{[C1]} Describe the interfaces that make up the core of the JDBC API, including the \texttt{Driver}, \texttt{Connection}, \texttt{Statement}, and \texttt{ResultSet} interfaces \newline
    \textbf{[C2]} Submit queries and read results from the database, including creating statements, returning result sets, iterating through the results, and properly closing result sets, statements, and connections
    \\ \midrule
    \textbf{[K18]} Localization KU&
    \textbf{[C1]} Read and set the locale by using the Locale object \newline
    \textbf{[C2]} Build a resource bundle for each locale and load a resource bundle in an application
    \\ \midrule
    \textbf{[K19]} Java Persistence KU&
    \textbf{[C1]} Create JPA Entity and Object-Relational Mappings (ORM) \newline
    \textbf{[C2]} Use Entity Manager to perform database operations, transactions, and locking with JPA entities \newline
    \textbf{[C3]} Create and execute JPQL statements \\ \midrule
    
    \textbf{[K20]} Enterprise Java Bean KU&
    \textbf{[S1]} Create session EJB components containing synchronous and asynchronous business methods, manage the life cycle container callbacks, and use interceptors. \newline
    \textbf{[S2]} Create EJB timers \\ \midrule

    \textbf{[K21]} Java Message Service API KU&
    \textbf{[S1]} Implement Java EE message producers and consumers, including Message-Driven beans \newline
    \textbf{[S2]} Use transactions with JMS API \\ \midrule
    
    \textbf{[K22]} SOAP Web Service KU&
    \textbf{[S1]} Create SOAP Web Services and Clients using JAX-WS API  \newline
    \textbf{[S2]} Create marshall and unmarshall Java Objects by using JAXB API \\ \midrule

    \textbf{[K23]} Servlet KU&
    \textbf{[S1]} Create Java Servlet and use HTTP methods \newline
    \textbf{[S2]} Handle HTTP headers, parameters, cookies \newline
    \textbf{[S3]} Manage servlet life cycle with container callback methods and WebFilters \\ \midrule

    \textbf{[K24]} Java REST API KU&
    \textbf{[S1]} Apply REST service conventions \newline
    \textbf{[S2]} Create REST Services and clients using JAX-RS API \\ \midrule

    \textbf{[K25]} Websocket KU&
    \textbf{[S1]} Create WebSocket Server and Client Endpoint Handlers \newline
    \textbf{[S3]} Produce and consume, encode and decode WebSocket messages \\ \midrule

    \textbf{[K26]} Java Server Faces KU&
    \textbf{[S1]} Use JSF syntax and use JSF Tag Libraries \newline
    \textbf{[S2]} Handle localisation and produce messages \newline
    \textbf{[S3]} Use Expression Language (EL) and interact with CDI beans \\ \midrule
    
    \textbf{[K27]} Contexts and Dependency Injection (CDI) KU&
    \textbf{[S1]} Create CDI Bean Qualifiers, Producers, Disposers, Interceptors, Events, and Stereotypes \\ \midrule 

    \textbf{[K28]} Batch Processing KU&
    \textbf{[S1]} Implement batch jobs using JSR 352 API \\
    \bottomrule
    \end{tabular}
   }
\end{table*}

\end{document}